\documentclass[showpacs,twocolumn,pre,floatfix]{revtex4}
\usepackage[dvips]{graphicx}
\usepackage[dvips]{color}
\newcommand{\href}[2]{#2}
\usepackage{amsmath}
\usepackage{amssymb}
\usepackage{bm}
\usepackage{times}
\allowdisplaybreaks
\newcommand{\vct}[1]{\mathbf{#1}}

\newcommand{\kap}{\mbox{\boldmath $\kappa$}}

\newcommand{\Om}{\Omega^\dagger}	
\newcommand{\struc}{\langle \varrho^*\varrho\rangle^{(\dot\gamma)}}
\newcommand{\struceq}{\langle \varrho^*\varrho\rangle}

\begin{document}
\title{
Nonequilibrium fluctuation dissipation relations of interacting  
Brownian particles driven by shear
}
\date{\today}
\author{Matthias Kr{\"u}ger and Matthias Fuchs}
\affiliation{Fachbereich Physik, Universit{\"a}t Konstanz, 78457 Konstanz, Germany}
\begin{abstract}
We present a detailed analysis of the fluctuation dissipation theorem (FDT) close to the glass transition in colloidal suspensions under steady shear using mode coupling approximations. Starting point is the many-particle Smoluchowski equation. Under shear, detailed balance is broken and the  response functions in the stationary state are smaller at long times than estimated from the equilibrium FDT. An asymptotically constant relation connects response and fluctuations during the shear driven decay, restoring the form of the FDT with, however, a ratio different from the equilibrium one. At short times, the equilibrium FDT holds. We follow two independent approaches whose results are in qualitative agreement.
To discuss the derived fluctuation dissipation ratios, we show an exact reformulation of the susceptibility which contains not the full Smoluchowski operator as in equilibrium, but only its well defined Hermitian part. This Hermitian part can be interpreted as governing the dynamics in the frame comoving with the probability current. 
We present a simple toy model which illustrates the FDT violation in the sheared colloidal system.
\end{abstract}


\maketitle
\section{Introduction}\label{sec:FDTFDR}
The thermal fluctuations of a system in equilibrium are directly connected to the system's response to a small external force. This connection, manifested in the fluctuation dissipation theorem (FDT), lies at the heart of linear response theory. The FDT {\it in equilibrium} connects the correlator $C_{\rm fg}^{(e)}(t)$ with the response function, the susceptibility $\chi_{\rm fg}^{(e)}(t)$ (both defined below), and reads 
\begin{equation}
\chi_{\rm fg}^{(e)}(t)=\frac{-1}{k_BT} \frac{\partial}{\partial t}C_{\rm fg}^{(e)}(t).\label{eq:FDTeq}
\end{equation} 
Eq.~\eqref{eq:FDTeq} states that the relaxation of a small fluctuation is independent of the origin of this fluctuation: Whether induced by a small external force or developed spontaneously by thermal fluctuations, the relaxation of the fluctuation cannot distinguish these cases.  
 
The most famous example for the FDT is the Einstein relation connecting the diffusivity of a Brownian particle to its mobility \cite{Einstein05}. 
The FDT is of importance for various applications in the field of material sciences since for example transport coefficients can be related to equilibrium quantities, i.e., the fluctuations of the corresponding variables \cite{Kubo}. It was first formulated by Nyquist in 1928 \cite{Nyquist28} as the connection between thermal fluctuations of the charges in a conductor (mean square voltage) and the conductivity. 

In non-equilibrium systems, this connection is not valid in general and much work is devoted to understanding the  relation between fluctuation and response functions. This relation is often characterized by the fluctuation dissipation ratio (FDR) $X_{\rm fg}(t)$ defined as
\begin{equation}
\chi_{\rm fg}(t) = - \frac{X_{\rm fg}(t)}{k_BT}\; \frac{\partial}{\partial t} C_{\rm fg}(t)\; .\label{eq:FDR}
\end{equation}
Close to equilibrium, one recovers the FDT in Eq.~\eqref{eq:FDTeq} with $X_{\rm fg}^{(e)}(t)\equiv1$. In non-equilibrium, $X_{\rm fg}(t)$ deviates from unity. This is related to the existence of non-vanishing probability currents (see Eq.~\eqref{eq:genFDT} below); FDRs are hence considered a possibility to quantify the currents and to signal non-equilibrium \cite{Crisanti03}. 
The violation of the equilibrium FDT has been studied for different systems before as we want to summarize briefly.

The general linear response susceptibility for non-equilibrium states with Fokker-Planck dynamics \cite{Risken} has been derived by Agarwal in 1972 \cite{agarwal72}. It will serve as exact starting point of our analysis (see Eq.~\eqref{eq:sus} below). The susceptibility is given in terms of microscopic quantities, which cannot easily be identified with a measurable function in general in contrast to the equilibrium case.  
For a single driven Brownian particle (colloid) in a periodic potential, the FDT violation for the velocity correlation has been studied in Ref.~\cite{Speck06}. 
There it was possible to compare the microscopic expressions successfully to the experimental realization of the system \cite{Blickle07}. 

Colloidal dispersions at high densities exhibit slow cooperative dynamics and form glasses. These metastable soft solids can be easily driven into stationary states far from equilibrium by already modest flow rates. Spin-glasses driven by non-conservative forces were predicted to exhibit nontrivial FDRs in mean field models \cite{Berthier99}. It is found that at long times the equilibrium form of the FDT (Eq.~\eqref{eq:FDTeq}) holds with the temperature $T$ replaced by a different value denoted  effective temperature $T_{\rm eff}$,
\begin{equation}
\hat\chi(\hat t)=\frac{-1}{k_BT_{\rm eff}} \frac{\partial}{\partial t}\hat C(\hat t).\label{eq:Berthier}
\end{equation}
This corresponds to a time independent FDR $\hat X_{\rm f}(\hat t)=\hat X_{\rm f}$ at long times during the final decay process, where $\hat t$ is time rescaled by the timescale of the external driving. In detailed computer simulations of binary Lennard-Jones mixtures by Berthier and Barrat \cite{Berthier02,Berthier02prl,Barrat00}, this restoration of the equilibrium FDT was indeed observed: For long times, the FDR $X_{\rm f}$ is independent of time. Its value was also very similar for the different investigated observables, i.e., $\hat X_{\rm f}=\hat X=T/T_{\rm eff}$ is proposed to be a universal number describing the non-equilibrium state. $T_{\rm eff}$ was found to be larger than the real temperature, which translates into an FDR smaller than unity. Further simulations with shear also saw $T_{\rm eff}>T$ \cite{Ohern04, Haxton07,Zamponi05,Ono02}, but the variable dependence was not studied in as much detail as in Ref.~\cite{Berthier02}, and partially other definitions of $T_{\rm eff}$ were used. In Refs.~\cite{Berthier99,Berthier02,Barrat02} it is argued that $T_{\rm eff}$ agrees with the effective temperature connected with the FDT violation in the corresponding aging system \cite{Barrat99, Kob99}. This has not yet been demonstrated for different temperatures. Note that the system under shear is always ergodic and aging effects are absent. The fluctuation dissipation relation of aging systems using mode coupling techniques was investigated in Ref.~\cite{Latz00}. Recently $T_{\rm eff}$ was also connected to barrier crossing rates \cite{Ilg07} replacing the real temperature in Kramers' escape problem \cite{Risken}. A theoretical approach for the effective temperature under shear in the so called ``shear-transformation-zone'' (STZ) model is proposed in Ref.~\cite{Langer07}.
Different techniques (with different findings) to measure FDRs in aging colloidal glasses were used in Refs.~\cite{Greinert06, Abou04, Maggi09}. No experimental realization of an FDT study of colloidal dispersions under shear is known to us.
An overview over the research situation (in 2003) can be found in Ref.~\cite{Crisanti03}.

Interesting universal FDRs were found in different spin models under coarsening \cite{Godreche00a, Godreche00b,Calabrese02,Mayer03,Sollich02} and under shear \cite{Corberi03}. At the critical temperature, a universal value of $X=\frac{1}{2}$ has been found e.g. in the $n$-vector-model for spatial dimension $d\geq 4$. In $d=3$, corrections to this value can dependent on the considered observable \cite{Calabrese04}. See Ref.~\cite{Calabrese05} for an overview. Yet, the situation for structural glasses has not been clarified. Also, the connection between structural glasses, spin glasses and critical systems is unclear.

In this paper, we present the study of the violation of the equilibrium FDT for dense colloidal suspensions under shear. It is a comprehensive extension of our recent Letter on the same topic \cite{Krueger09}, but also provides a number of new results and discussions. We build on the MCT-ITT approach \cite{Fuchs02,Fuchs03,Fuchs05,Fuchs09} (reviewed recently \cite{Fuchs08b}) based on mode coupling theory. This approach allows us to derive quantities which are directly measured in experiments and simulations \cite{Besseling07,Zausch08,Varnik08} and the properties of specific observables can be described. We will hence be able to study the non-equilibrium FDT for different observables as measured in simulations, and possible differences for different variables can be detected. In the main text, we will follow the calculation as presented in Ref.~\cite{Krueger09} in detail. It leads to a time independent FDR $\hat X_{\rm f}$ during the whole final relaxation process whose value is universal in the simplest approximation, $\hat X_{\rm f}=\frac{1}{2}$. We will also derive corrections to this value which depend on the considered observable. In Appendix \ref{sec:ZM}, we will additionally show a different analysis of the extra term in the FDT following more standard routes of MCT and projection operator formalisms. It is in qualitative agreement with the results shown in the main text and it allows us to estimate the size of the correction terms which are neglected in the main text and to see that they are small.

The paper is organized as follows. In Sec.~\ref{sec:susc}, we will introduce the microscopic starting point and give the definitions of the different time dependent correlation and response functions. In Sec.~\ref{sec:viol}, we will introduce the different contributions to the non-equilibrium term $\Delta\chi_{\rm fg}(t)$ in the susceptibility. These different contributions are approximated in Sec.~\ref{sec:approx}. The approximations for the time dependent correlation functions will be shown in Sec.~\ref{sec:twotime}. In Sec.~\ref{sec:FDres}, we will present the final extended FDT connecting the susceptibility to measurable correlation functions and discuss the FDR as function of different parameters. In Sec.~\ref{sec:Corr}, we show an exact form of the susceptibility which involves the Hermitian part of the Smoluchowski operator and the restoration of the equilibrium FDT in the frame comoving with the probability current. The final discussion, supported by the FDR analysis in a simple toy model will finally be presented in Sec.~\ref{sec:disc}. In Appendix \ref{sec:ZM}, we derive the expressions for the susceptibility in an approach based on the Zwanzig-Mori projection operator formalism.
\section{Microscopic starting point}\label{sec:susc}
We consider a system of $N$ spherical Brownian particles of diameter $d$, dispersed in a solvent. The system has volume $V$. The particles have bare diffusion constants $D_0$. The interparticle force acting on particle $i$ ($i=1\dots N$) at position $\vct{r}_i$ is given by $\vct{F}_i=-\partial/\partial \vct{r}_i U(\{\vct{r}_j\})$, where $U$ is the total potential energy. We neglect hydrodynamic interactions to keep the description as simple as possible. These are also absent in the computer simulations \cite{Berthier02} to which we will compare our results. 

The external driving, viz. the shear, acts on the particles via the solvent flow velocity $\vct{v}(\vct{r})=\dot\gamma y \hat{\vct{x}}$, i.e., the flow points in $x$-direction and varies in $y$-direction. $\dot\gamma$ is the shear rate. The particle distribution function $\Psi(\Gamma\equiv\{{\vct{r}_i}\},t)$ obeys the Smoluchowski equation \cite{Dhont,Fuchs05},
\begin{eqnarray}\label{eq:smol}
\partial_t \Psi(\Gamma,t)&=&\Omega \; \Psi(\Gamma,t),\nonumber\\
\Omega&=&\Omega_e+\delta\Omega=\sum_{i}\boldsymbol{\partial}_i\cdot\left[\boldsymbol{\partial}_i-{\bf F}_i - \kap\cdot\vct{r}_i\right],
\end{eqnarray}
with $\kap=\dot\gamma\hat{\vct{x}}\hat{\vct{y}}$ for the case of simple shear.
$\Omega$ is called the Smoluchowski operator (SO) and it is built up by the equilibrium SO, $\Omega_e=\sum_{i}\boldsymbol{\partial}_i\cdot[\boldsymbol{\partial}_i-{\bf F}_i]$ of the system without shear and the shear term $\delta\Omega=-\sum_i\boldsymbol{\partial}_i\cdot \kap\cdot\vct{r}_i$. We introduced dimensionless units for space, energy and time, $d=k_BT=D_0=1$. The formal H-theorem \cite{Risken} states that the system reaches the equilibrium distribution $\Psi_e$, i.e., $\Omega_e \Psi_e=0$, without shear. Under shear, the system reaches the stationary distribution $\Psi_s$ with $\Omega \Psi_s=0$. Ensemble averages in equilibrium and in the stationary state are denoted
\begin{subequations}
\begin{eqnarray}
\left\langle\dots\right\rangle&=&\int d\Gamma \Psi_e(\Gamma) \dots,\\
\left\langle\dots\right\rangle^{(\dot\gamma)}&=&\int d\Gamma \Psi_s(\Gamma) \dots ,
\end{eqnarray}
\end{subequations}
respectively.
 In the stationary state, the distribution function is constant but the system is not in thermal equilibrium due to the non-vanishing probability current $\vct{j}_i^s$ \cite{Fuchs05},
\begin{equation}
\vct{j}_i^s=[-\boldsymbol{\partial}_i+{\bf F}_i +\kap\cdot\vct{r}_i]\Psi_s\label{eq:current}={\boldsymbol{\hat\jmath}}_i\Psi_s.
\end{equation} 
\subsection{Correlation functions}\label{sec:correlation}
\begin{figure}
\centering{\includegraphics[width=0.6\linewidth]{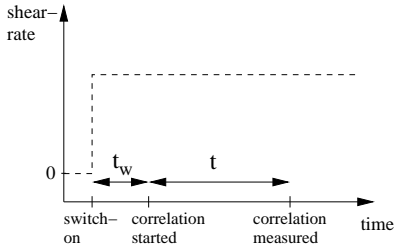}}
\caption{\label{fig:waiting}Definition of the waiting time $t_w$ and the correlation time $t$ after switch-on of the rheometer.}
\end{figure}

Dynamical properties of the system are probed by time dependent correlation functions. The correlation of the fluctuation $\delta f=f-\langle f \rangle^{(\dot\gamma)}$ of a function $f(\{\vct{r}_i\})$ with the fluctuation of a function  $g(\{\vct{r}_i\})$ is in the stationary state given by \cite{Fuchs05}
\begin{equation}
C_{\rm fg}(t)=\left\langle\delta f^* e^{\Omega^\dagger t} \delta g\right\rangle^{(\dot\gamma)}.\label{eq:stationary}
\end{equation}
Here, $\Omega^\dagger=\sum_i[\boldsymbol{\partial}_i+\vct{F}_i+\vct{r}_i\cdot\kap^T]\cdot\boldsymbol{\partial}_i$ is the adjoint SO that arose from partial integrations.
$C_{\rm fg}(t)$ is called the stationary correlator, it is the correlation function which is mostly considered in experiments and simulations of sheared suspensions. At this point, we would like to introduce three more correlation functions which will appear in this paper. The transient correlator $C_{\rm fg}^{(t)}$ is observed when the external shear is switched on at $t=0$ \cite{Fuchs09}, 
\begin{equation}
C^{(t)}_{\rm fg}(t)=\left\langle \delta f^*e^{\Omega^\dagger t}\delta g\right\rangle.\label{eq:tran}
\end{equation}
It probes the dynamics in the transition from equilibrium to steady state and is the central object of the MCT-ITT approach \cite{Fuchs03,Fuchs09}. 
In the general case, where the correlation is started a period $t_w$, namely the waiting time, after the rheometer was switched on, one observes the two-time correlator $C_{\rm fg}(t,t_w)$, see Fig.~\ref{fig:waiting}, 
\begin{equation}
C_{\rm fg}(t,t_w)=C_{\rm fg}^{(t)}(t)+\dot\gamma\int_0^{t_w} ds\left\langle\sigma_{xy}e^{\Omega^\dagger s} \delta f^{*}e^{\Omega^{\dagger} t}  \delta g \right\rangle.\label{eq:2ti}
\end{equation}
Eq.~\eqref{eq:2ti} follows with the waiting time dependent distribution function \cite{Fuchs05},
\begin{equation}
\Psi(\Gamma,t_w)=\Psi_e+\int_0^{t_w} ds\,e^{\Omega s}\Omega \Psi_e,\label{eq:psiwai}
\end{equation}
with partial integrations when averaging with $\Psi(\Gamma,t_w)$. For $t_w=0$, the two-time correlator equals the transient correlator. For long waiting times, it reaches the stationary correlator, $C_{\rm fg}(t,t_w\to\infty)\to C_{\rm fg}(t)$, and Eq.~\eqref{eq:2ti} becomes the ITT expression for $C_{\rm fg}(t)$ \cite{Fuchs05}. Without shear finally, one observes the equilibrium correlation,
\begin{equation}
C_{\rm fg}^{(e)}=\left\langle\delta f^* e^{\Omega_e^\dagger t} \delta g\right\rangle.
\end{equation}
\subsection{Response Functions}
The susceptibility $\chi_{\rm fg}$ describes the linear response of the stationary system to an external perturbation. Note that the term `linear response' does not correspond to the shear, but to the additional small test force $h_e(t)$ acting on the particles. Because the system is always ergodic due to shearing, the linear response will always exist in contrast to un-sheared glasses \cite{Gazuz08,Habdas04}, where a finite force is needed to mobilize the particles. Formally, the susceptibility $\chi_{\rm fg}(t)$ describes the linear response of the stationary expectation value of $g$ to the external perturbation $h_e(t)$ which  shifts the internal energy $U$ to $U-f^*\,h_e(t)$,
\begin{equation}
\left\langle g \right\rangle^{(\dot\gamma,h_e)}(t)-\left\langle g
\right\rangle^{(\dot\gamma)}=\int_{-\infty}^t \!\!\!\!\!dt'\chi_{\rm fg}(t-t')h_e(t')+\mathcal{O}(h_e^2)\; .\label{eq:defsus}
\end{equation}
To derive the microscopic form of the susceptibility, one considers the change of the stationary distribution function $\Psi_s$ under the external perturbation. One finds \cite{agarwal72, Fuchs05,Risken} 
\begin{equation}
\chi_{\rm fg}(t)=\left\langle \sum_i \frac{\partial  f^*}{\partial {\bf r}_i}\cdot\boldsymbol{\partial}_i e^{\Omega^\dagger t} g\right\rangle^{(\dot\gamma)}.\label{eq:sus}
\end{equation}
If one replaces $\Psi_s$ by $\Psi_e$ and $\Omega^\dagger$ by $\Omega^\dagger_e$ in Eq.~\eqref{eq:sus}, the equilibrium FDT in Eq.~\eqref{eq:FDTeq} follows by partial integrations,
\begin{align}
\chi^{(e)}_{\rm fg}(t)=&\left\langle \sum_i \frac{\partial  f^*}{\partial {\bf r}_i}\cdot\boldsymbol{\partial}_i e^{\Omega_e^\dagger t} g\right\rangle\nonumber\\=&-\left\langle f^*\Omega^\dagger_e e^{\Omega_e^\dagger t} g\right\rangle=-\frac{\partial}{\partial t} C_{\rm fg}^{(e)}(t). 
\end{align}
In the considered non-equilibrium system, where
detailed balance is broken and the nonzero stationary probability
current in Eq.~\eqref{eq:current} exists, the equilibrium FDT \eqref{eq:FDTeq} is extended as we see now. The above expression, Eq.~\eqref{eq:sus}, can be rewritten (with
adjoint current operator $\boldsymbol{\hat\jmath}_i^\dagger=\boldsymbol{\partial}_i+\vct{F}_i+\vct{r}_i\cdot\kap^T$) to
\begin{equation}
\Delta\chi_{\rm fg}(t)=\chi_{\rm fg}(t)+\dot C_{\rm fg}(t)=-\left\langle\sum\limits_{
i}\boldsymbol{\hat\jmath}_i^\dagger\cdot\frac{\partial f^*}{\partial\mathbf{r}_i
}e^{\Omega^{\dagger} t}g \right\rangle^{(\dot\gamma)}\label{eq:genFDT}.
\end{equation}
Note that the new term in the FDT, $\Delta\chi_{\rm fg}(t)$, in the following called {\it violating term}, is directly proportional to the stationary probability current.
A deviation of the fluctuation dissipation ratio, Eq.~\eqref{eq:FDR},
from unity, the value close to equilibrium, arises. 

The extended FDT in Eq.~(\ref{eq:genFDT}) has been known since the work
of Agarwal \cite{agarwal72}. We will analyze it for driven
metastable (glassy) states and show that the additive
correction $\Delta\chi_{\rm fg}(t)$ \cite{Blickle07, Speck06, Harada05} leads to the nontrivial constant FDR at long  times, as was found in the simulations.  One can always express the FDT violation in terms of an additive as well as a multiplicative correction. The nontrivial statement for driven metastable glasses is that the multiplicative correction is possible with a time independent factor at long times. Specifically, we will
look at autocorrelations, $g=f$ of functions
 without explicit advection,
$f=f(\{y_i,z_i\})$, where the flow-term in the
current operator $\boldsymbol{\hat\jmath}_i^\dagger$ in (\ref{eq:genFDT}) vanishes. For variables depending on $x_i$, the equilibrium FDT is already violated for low colloid densities as seen from the Einstein relation: The mean squared displacement grows cubically in time \cite{Elrick62} (Taylor dispersion), while the mobility of the particle is constant.

In contrast to the equilibrium distribution, $\Psi_e\propto e^{-U}$, the stationary distribution is not known and stationary averages are calculated in the integration through transients approach (ITT) \cite{Fuchs05} (compare Eq.~\eqref{eq:psiwai}),
\begin{equation}
\langle \dots\rangle^{(\dot\gamma)}=\langle
\dots\rangle+\dot\gamma\int_0^\infty
ds\langle\sigma_{xy}e^{\Omega^\dagger s}\dots\rangle\; .\nonumber
\end{equation}
ITT simplifies the following
analysis because averages can now be evaluated in equilibrium, while
otherwise non-equilibrium forces would be required \cite{Szamel}.
E.g.~due to $\boldsymbol{\partial}_i \Psi_e={\bf F}_i \Psi_e$, the
expression (\ref{eq:genFDT}) vanishes in the equilibrium average and reduces to
\begin{equation}
\Delta\chi_{\rm f}(t)=-\dot\gamma\int_0^\infty ds \left\langle\sigma_{xy}e^{\Omega^{\dagger} s}\sum\limits_{i}(\boldsymbol{\partial}_i+{\bf F}_i)\cdot\frac{\partial f^*}{\partial \vct{r}_i}e^{\Omega^{\dagger} t}f\right\rangle.
\label{eq:deltachi}
\end{equation}
Eq.~\eqref{eq:deltachi} is still exact and we will in the following develop approximations for it. In the main text, we will follow the derivation as presented in Ref.~\cite{Krueger09}. In  Appendix \ref{sec:ZM}, we present an alternative derivation with Zwanzig Mori projections. We will show in Sec.~\ref{sec:quant} that the two approaches are in qualitative agreement. 
\section{The violating term}\label{sec:viol} 
We  want to analyze the violating term $\Delta\chi_{\rm f}(t)$ in more detail. It can be split up into terms containing the Smoluchowski operator instead of the unfamiliar operator $\boldsymbol{\hat\jmath}_i^\dagger$. This  can be done with the following identity for general functions $f(\{{\bf r}_i\})$ and $g(\{{\bf r}_i\})$,
\begin{equation}
\sum\limits_{i}\boldsymbol{\hat\jmath}_i^\dagger\cdot\frac{\partial f^*}{\partial\mathbf{r}_i} g=\frac{1}{2} \left[\Omega^\dagger f^*g-f^*\Omega^\dagger g+(\Omega^\dagger f^*)g\right].\label{eq:sep}
\end{equation} 
If we apply this identity to Eq.~\eqref{eq:deltachi} with $g=e^{\Om t}f$, we get the following three terms,
\begin{align}
&\Delta\chi_{\rm f}(t)=\nonumber\\&\frac{-\dot\gamma}{2}\int_0^\infty ds \bigl\langle \sigma_{xy}e^{\Omega^\dagger s} [\textcolor{black}{\Omega^\dagger}  f^*-f^*\textcolor{black}{\Omega^\dagger}
+(\Omega^\dagger f^*)]e^{\Omega^\dagger t}f\bigr\rangle\label{eq:dom}.
\end{align}
The first term in Eq.~\eqref{eq:dom} contains a derivative with respect to $s$ and the $s$-integration can immediately be done. We find that the first term in Eq.~\eqref{eq:dom} (without the factor $\frac{1}{2}$) exactly describes the derivative of $C_{\rm f}(t,t_w)$ with respect to $t_w$ at $t_w=0$,
\begin{align}
&-\dot\gamma\int_0^\infty ds \left\langle \sigma_{xy}e^{\Omega^\dagger s} \Omega^\dagger \delta f^*e^{\Omega^\dagger t}\delta f\right\rangle=
\dot\gamma \left\langle \sigma_{xy}  \delta f^*e^{\Omega^\dagger t} \delta f\right\rangle\nonumber\\
&=\left\langle \delta f^* \delta\Omega^\dagger e^{\Omega^\dagger t}\delta f\right\rangle =\left.\frac{\partial}{\partial t_w} C_{\rm f}(t,t_w)\right|_{t_w=0},
\label{eq:chi}
\end{align}
where from now on, we consider fluctuations from equilibrium, $\delta f=f- \langle f\rangle$. The constant $\langle f\rangle$ cancels in (\ref{eq:dom}). The second equal sign in Eq.~\eqref{eq:chi} follows with partial integrations (recall $\delta\Omega\Psi_e=\sigma_{xy}\Psi_e$ and $[\delta \Omega^\dagger,\delta f]=0$). The intriguing connection to the waiting time derivative follows by comparison of the right hand side of the first line of Eq.~\eqref{eq:chi} with Eq.~\eqref{eq:2ti}.

 The second term in Eq.~\eqref{eq:dom} describes the time derivative of the difference between stationary and transient correlator, compare Eq.~\eqref{eq:2ti} with $t_w\to\infty$. The last term in Eq.~\eqref{eq:dom} has no physical interpretation and we denote it by $\Delta\chi^{(3)}_{\rm f}$. We hence have
\begin{align}
\Delta\chi_{\rm f}(t)=&\frac{1}{2}\left.\frac{\partial}{\partial t_w} C_{\rm f}(t,t_w)\right|_{t_w=0}\nonumber\\&+ \frac{1}{2}\left[\dot C_{\rm f}(t)-\dot C^{(t)}_{\rm f}(t)\right]+\Delta\chi^{(3)}_{\rm f}\,,
\end{align}
where $\Delta\chi^{(3)}_{\rm f}=\frac{-\dot\gamma}{2}\int_0^\infty ds \langle \sigma_{xy}e^{\Omega^\dagger s} (\Omega^\dagger f^*)e^{\Omega^\dagger t}f\rangle$.
In the following subsections, we will look at the different terms more closely. 
\section{Approximations for the violating term}\label{sec:approx}
\subsection{The waiting time derivative}
In order to approximate the waiting time derivative in Eq.~\eqref{eq:chi}, we note its connection to time derivatives of correlation functions,
\begin{equation}
\left\langle  \delta f^* \,\delta\Omega^\dagger \,e^{\Omega^\dagger t} \delta f\right\rangle=\dot C_{\rm f}^{(t)}(t)-\left\langle  \delta f^* \,\Omega^\dagger_e\,e^{\Omega^\dagger t} \delta f\right\rangle\label{eq:ableitungen}.
\end{equation}
The time derivative of the transient correlator $C_{\rm f}^{(t)}(t)$ is split into two terms, one containing the equilibrium operator $\Omega^\dagger_e$, the other one containing the shear term $\delta\Omega^\dagger$. We will reason the following: The term containing $\Omega^\dagger_e$ is the derivative of the short time, shear independent dynamics of the transient correlator down on the plateau (compare Fig.~\ref{fig:twotimefull} below), i.e., the derivative of the dynamics governed by the equilibrium SO $\Omega_e$. The term containing $\delta\Omega^\dagger$, i.e., the waiting time derivative, follows then as the time derivative with respect to the shear governed decay from the plateau down to zero.

The equilibrium derivative $\Omega^{\dagger}_e \delta f^*$ in the last term of \eqref{eq:ableitungen} de-correlates quickly as the particles loose memory of their initial motion even without shear. In this case, the latter term is the time derivative of the equilibrium correlator, $C^{(e)}_{\rm f}(t)$. A shear flow switched on at $t=0$ should make the particles forget their initial motion even faster, prompting us to use the approximation $e^{\Omega^{\dagger} t} \approx  e^{\Omega^{\dagger}_e t} P_{\rm f} e^{-\Omega^{\dagger}_e t}\, e^{\Omega^{\dagger}t}$, with projector $P_{\rm f}=\delta f^*\rangle\langle\delta f^*\delta f\rangle^{-1}\langle\delta f$. We then find
\begin{equation}
\left\langle \delta f^* \Omega^\dagger_ee^{\Omega^\dagger t}\delta f\right\rangle\approx\left\langle \delta f^* \Omega^\dagger_ee^{\Omega^\dagger_e t}\delta f\right\rangle\frac{\left\langle \delta f^* e^{-\Omega^\dagger_e t} e^{\Omega^\dagger t}\delta f\right\rangle}{\langle \delta f^* \delta f\rangle}.\label{eq:approx1}
\end{equation}
The first average on the right hand side is the time derivative of $C^{(e)}_{\rm f}(t)$. The second average is not known. Applying the same approximation $e^{\Omega^{\dagger} t} \approx  e^{\Omega^{\dagger}_e t} P_{\rm f} e^{-\Omega^{\dagger}_e t}\, e^{\Omega^{\dagger}t}$ to the transient correlator, we have
\begin{equation}
\left\langle \delta f^* e^{\Omega^\dagger t}\delta f\right\rangle\approx\left\langle \delta f^* e^{\Omega^\dagger_e t}\delta f\right\rangle\frac{\left\langle \delta f^* e^{-\Omega^\dagger_e t} e^{\Omega^\dagger t}\delta f\right\rangle}{\langle \delta f^* \delta f\rangle}.
\end{equation}
Combining the two equations, we find for the last term in Eq.~\eqref{eq:ableitungen}
\begin{equation}
\left\langle \delta f^* \Omega^\dagger_ee^{\Omega^\dagger t}\delta f\right\rangle\approx\dot C^{(e)}_{\rm f}(t)\frac{C^{(t)}_{\rm f}(t)}{C^{(e)}_{\rm f}(t)}.\label{eq:fast}
\end{equation}
This term is then assured to decay faster than without shear. Now we can give the final formula for the waiting time derivative,
\begin{equation}
\left.\frac{\partial}{\partial t_w} C_{\rm f}(t,t_w)\right|_{t_w=0} \approx \dot C^{(t)}_{\rm f}(t)-\dot C^{(e)}_{\rm f}(t)\frac{C^{(t)}_{\rm f}(t)}{C^{(e)}_{\rm f}(t)}\label{eq:lti}.
\end{equation}
This is our central approximation whose consequences for the FDR will be worked out in Sec.~\ref{sec:FDres}. The quality of approximation \eqref{eq:lti} has  recently been studied in detailed simulations, and qualitative and quantitative agreement was found for two different simulated super-cooled liquids \cite{twotime}. 
As argued above, the last term in (\ref{eq:lti}) will be identified as {\it short
time derivative} of $C^{(t)}_{\rm f}$, connected with the shear
independent decay, where the transient correlator equals the equilibrium correlator. Consequently, $\frac{\partial}{\partial t_w} C_{\rm f}(t,t_w)|_{t_w=0}$ will turn out to be the {\it long time
derivative} of $C^{(t)}_{\rm f}$, connected with the final shear driven
decay. This captures the additional
dissipation provided by the coupling to the stationary probability
current in Eq.~(\ref{eq:genFDT}). The approximation in Eq.~\eqref{eq:lti} is also reasonable comparing it to the expected properties of the waiting time derivative: For long times, $t\to\infty$ and $\dot\gamma\to0$ with $\dot\gamma t=\mathcal{O}(1)$, one has $\dot C^{(e)}_{\rm f}(t)=0$ in glassy states and the waiting time derivative is equal to the time derivative of the transient correlator. Varying the waiting time or the correlation time has then the same effect on the transient correlator. It is for small waiting times a function of $\dot\gamma(t+t_w)$ since $(t+t_w)$ measures the time since switch-on \cite{Krueger09b}.
\subsection{The other terms in Eq.~\eqref{eq:dom}}\label{sec:other}
The second term in Eq.~\eqref{eq:dom} has a physical interpretation as well: It is the time derivative of the difference between stationary and transient correlator, see Eq.~\eqref{eq:2ti},
\begin{equation}
\dot\gamma\int_0^\infty ds \langle\sigma_{xy}e^{\Omega^\dagger s} \delta f^*\Omega^\dagger e^{\Omega^\dagger t}\delta f\rangle=\dot C_{\rm f}(t)-\dot C_{\rm f}^{(t)}(t).\label{eq:zweiter}
\end{equation} 
The last term, $\Delta\chi^{(3)}_{\rm f}$, has yet no physical interpretation. At $t=0$, it cancels with the second term. It is a demanding task to estimate the contribution of the different terms to $\Delta\chi_{\rm f}(t)$. This can be done in an MCT analysis for density fluctuations as presented in Appendix \ref{sec:ZM}. We want to briefly summarize the results for the contributions of the different terms as found in Appendix \ref{sec:ZM}. For coherent, i.e., collective density fluctuations, we have $f=\varrho_{\bf q}=\sum_i e^{i\,{\bf q}\cdot{\bf r}_i}$ \cite{Hansen}. For incoherent, ie., single particle fluctuations, one has $f=\varrho_{\bf q}^s=e^{i\,{\bf q}\cdot{\bf r}_s}$ with ${\bf r}_s$ the position of the tagged particle. We denote all normalized density functions with subscript ${\bf q}$, the normalized transient density correlator is denoted $\Phi_{\bf q}(t)$ \cite{Fuchs03}. We find that the violating term is zero for $\dot\gamma t\ll1$ \cite{footnote1}. For long times, $\dot\gamma t=\mathcal{O}(1)$, we can estimate the different contributions in terms of $\Phi_{\bf q}(t)$, as is shown in Appendix \ref{sec:ZM}. We find
\begin{subequations}\label{eq:estimate}
\begin{eqnarray}
\hspace{-1cm}\frac{1}{2}\left.\frac{\partial}{\partial t_w} C_{\bf q} (t,t_w)\right|_{t_w=0}&\approx& \tilde n_{\bf q}^{(1)} \Phi_{\bf q}(t),\\
\frac{1}{2}\left[\dot C_{\bf q}(t)-\dot \Phi_{\bf q}(t)\right]+\Delta\chi^{(3)}_{\bf q}(t)&\approx& \tilde n_{\bf q}^{(2+3)} \Phi_{\bf q}(t),
\end{eqnarray} 
\end{subequations}
\begin{figure}
\begin{center}\includegraphics[width=1\linewidth]{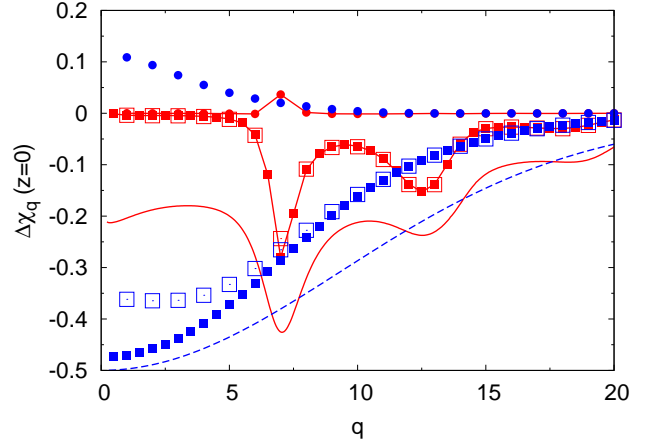}\end{center}
\caption{\label{fig:memories}The estimates for the Laplace transform $\Delta\chi_{\bf q}(z=0)$ of the violating term in density susceptibilities for ${\bf q}=q\hat{\bf y}$. We show the coherent (incoherent) case with data points connected (unconnected) by lines. Solid squares show the estimate for the first term, the waiting time derivative, solid spheres show the sum of the two other terms. Open squares show the sum of all three. The solid (dashed) line shows the prediction of Eq.~\eqref{eq:lti} for the waiting time derivative for the coherent (incoherent) case; compare with the solid squares and see main text. The picture is very similar for the $z$-direction \cite{Krueger09b}.}
\end{figure}
where the functions $\tilde n_{\bf q}\propto |\dot\gamma|$ for small shear rates in glassy states.
In Fig.~\ref{fig:memories}, we show the Laplace transform $\mathcal{L}\{\Delta\chi_{\bf q}(t)\}(z)=\int_0^\infty dt e^{-zt} \Delta\chi_{\bf q}(t)$ at $z=0$ for the different contributions to $\Delta\chi_{\bf q}$ as estimated in Appendix \ref{sec:ZM}. We see that, according to the estimates in Appendix \ref{sec:ZM}, the first term in Eq.~\eqref{eq:dom} is the dominant contribution to the violating term. It is larger than the two other terms, which additionally partially cancel each other. We will in the main text neglect the sum of second and third term, this will give good agreement to the data in Ref.~\cite{Berthier02}. In Fig.~\ref{fig:memories}, we also show the prediction of Eq.~\eqref{eq:lti} for the waiting time derivative, which is at $z=0$ simply given by minus the height of the glassy plateau, see Eq.~\eqref{eq:waitz0}. We see that our estimate in Appendix \ref{sec:ZM} agrees qualitatively and also semi-quantitatively with Eq.~\eqref{eq:lti}. 

It appears reasonable to conclude that the waiting time derivative is larger than e.g. the second term \eqref{eq:zweiter}, since it is equal to the time derivative of the transient correlator  at long times, whereas \eqref{eq:zweiter} is the difference of two very similar functions. It is equal to the third term at $t=0$ which lets us expect that also $\Delta\chi_{\rm f}^{(3)}$ is small.
\section{Approximations for Correlation functions}\label{sec:twotime}
\subsection{ITT equations for transient correlators}
The known ITT solutions for the transient correlators will be the central input for our FDR analysis. In order to visualize its time dependence, we will use the schematic $F_{12}^{(\dot\gamma)}$-model of ITT, which has
repeatedly been used to investigate the dynamics of quiescent and
sheared dispersions \cite{Fuchs03}, and which provides excellent
fits to the flow curves from large scale simulations
\cite{Varnik06a}. It provides a normalized transient correlator
$C^{(t)}(t)=\Phi(t)$, as well as a quiescent one,
representing coherent, i.e., collective density fluctuations. The equation of motion reads \cite{Fuchs03},
\begin{subequations}
\label{eq:f12}
\begin{eqnarray}
0=\dot\Phi(t)+\Gamma\left\{\Phi(t)+\int_0^t dt'm(\dot\gamma,t-t')\dot\Phi(t')\right\},\\
m(\dot\gamma,t)=\frac{1}{1+(\dot\gamma
t)^2}\left[(v_1^c+2.41\varepsilon)\Phi(t)+v^c_2\Phi^2(t)\right],\label{eq:f122}
\end{eqnarray}
\end{subequations}
with initial decay rate $\Gamma$. We use the much studied
values $v^c_2=2$, $v^c_1=v^{c}_2(\sqrt{4/v^{c}_2}-1)$ with the glass
transition at $\varepsilon=0$, and take $m(0,t)$ in order to
calculate quiescent ($\dot\gamma=0$) correlators \cite{Goetze84}. Positive values of the separation parameter $\varepsilon$ correspond to glassy states, negative values to liquid states. 
 
In order to study the $q$-dependence of our results, we will use the isotropic approximation \cite{Fuchs03} for the normalized transient density correlator. For glassy states, the final decay from the glassy plateau of height $f_q$ is approximated as exponential, 
\begin{equation}
\Phi_{\bf q}(t)\approx\Phi_q(t)=f_q\,e^{-c\frac{h_q}{f_q} |\dot\gamma| t}\label{eq:corr1},
\end{equation} 
where the amplitude $h_q$ is also derived within quiescent MCT \cite{Goetze91}.
\subsection{Two-time and stationary correlator}
\begin{figure}
\begin{center}\includegraphics[width=1\linewidth]{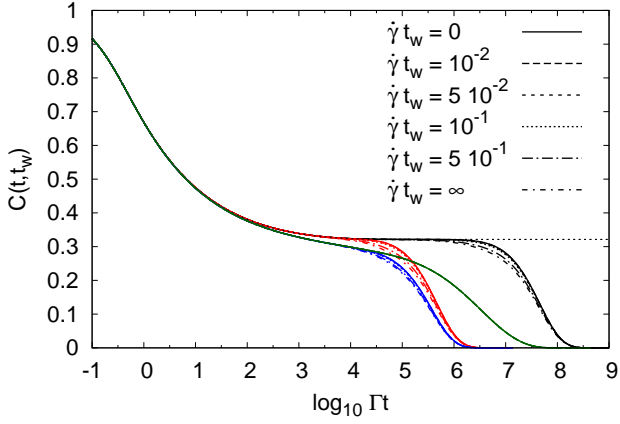}\end{center}  
\caption{$C(t,t_w)$ from Eq.~\eqref{eq:stat1} for different waiting times as indicated by the different line styles. The $F_{12}^{(\dot\gamma)}$ model Eq.~\eqref{eq:f12} is used to generate the transient correlators $C^{(t)}$ (solid lines). We show a glassy state ($\varepsilon=10^{-3}$) at shear rates $\dot\gamma/\Gamma=10^{-8}$ (black) and $\dot\gamma/\Gamma=10^{-6}$ (red). Green and blue curves show the same respective shear rates for a liquid state ($\varepsilon=-10^{-3}$). The thin dashed line shows the equilibrium correlator $C^{(e)}(t)$ for the glassy state. For the liquid state, the equilibrium correlator coincides with the green curve.}
\label{fig:twotimefull}                                                         \end{figure}

We will need to know the difference between stationary and transient correlators in order to be able to study the FDR in detail. Here we derive an approximate expression for the two-time correlator $C_{\rm f}(t,t_w)$, which then gives the stationary correlator for $t_w\to\infty$. The detailed discussion will be presented elsewhere \cite{twotime,Krueger09b}. We start from the exact Eq.~\eqref{eq:2ti} and use the projector $\sigma_{xy}\rangle\langle\sigma_{xy}\sigma_{xy}\rangle^{-1}\langle\sigma_{xy}$ as well as Eq.~\eqref{eq:chi} to get
\begin{equation}
C_{\rm f}(t,t_w)-C^{(t)}_{\rm f}(t)\approx\int_0^{t_w} ds\frac{\langle\sigma_{xy}e^{\Omega^{\dagger} s}\sigma_{xy}\rangle}{\left\langle\sigma_{xy}\sigma_{xy}\right\rangle}\left.\frac{\partial C_{\rm f}(t,t_w)}{\partial t_w}\right|_{t_w=0}.
\label{eq:stat1}
\end{equation}
Eq.~\eqref{eq:stat1} is a short version  which neglects the waiting time dependence of the $t=0$ value of the unnormalized two-time correlator. An extended version including this effect can be formulated \cite{twotime,Krueger09b} but it is  more involved and would change our results only marginally. Thus, we continue with Eq.~\eqref{eq:stat1}  where the first factor on the right hand side is the normalized integrated shear modulus 
\begin{equation}
\tilde\sigma(t_w)\equiv\dot\gamma\int_0^{t_w} ds\frac{\langle\sigma_{xy}e^{ \Omega^{\dagger} s}\sigma_{xy}\rangle}{\langle\sigma_{xy}\sigma_{xy}\rangle},
\end{equation}
containing as numerator the familiar stationary shear stress, measured in 'flow curves' as function of shear rate \cite{Fuchs05,Crassous08,Siebenbuerger09}. A technical problem arises for hard spheres, where the instantaneous shear modulus $\langle\sigma_{xy}\sigma_{xy}\rangle$ diverges \cite{Fuchs03} giving formally $\tilde\sigma=0$. The proper limit of increasing steepness in the repulsion has to be addressed in the future \cite{twotime}. 
In the spirit of the $F_{12}^{(\dot\gamma)}$-model \cite{footnote2}, we approximate the $s$-dependent normalized shear modulus by the transient correlator \cite{Fuchs03,Krueger09},
\begin{equation}
\frac{\langle\sigma_{xy}e^{\Omega^{\dagger} s}\sigma_{xy}\rangle}{\langle\sigma_{xy}\sigma_{xy}\rangle}\approx \Phi(s)\frac{G_\infty}{f}\label{eq:mod},
\end{equation}
where we account for the different plateau heights of the respective normalized functions by setting $G_\infty/f\approx\frac{1}{3}$; choosing a quadratic dependence $\Phi^2(s)$ would only change the results imperceptibly. We will abbreviate $\tilde\sigma(t_w\to\infty)=\tilde\sigma$. 
The second factor on the right of Eq.~\eqref{eq:stat1} is the waiting time derivative, which we approximated in Eq.~\eqref{eq:lti}. We are hence able to show the two-time correlator for different waiting times for a glassy and a liquid state (Fig.~\ref{fig:twotimefull}). The short time decay down onto the plateau $f$ is independent of waiting time $t_w$, whereas the long time decay becomes slightly faster with increasing waiting time. Overall the waiting-time dependence is small.
In recent simulations of density fluctuations of
soft spheres \cite{Zausch08}, the difference between the two
correlators
 was found to be largest at intermediate
times, and $C_{\rm f}(t,t_w)\leq C_{\rm f}^{(t)}(t)$ was observed. Both properties
are fulfilled by Eq.~(\ref{eq:stat1}). Note that Eq.~(\ref{eq:stat1}) is exact in first order in $t_w$.

Based on Fig.~\ref{fig:twotimefull} and the knowledge about the transient correlators \cite{Fuchs03,Fuchs09}, the short time decay of $C(t,t_w)$ is independent of  shearing for small shear rates. In glassy states at $\dot\gamma\to0$ with $\dot\gamma t=const.$, the transient correlator reaches a scaling function $\hat C^{(t)}_{\rm f}(\dot\gamma t)$ \cite{Fuchs03}, and the two-time correlator from Eq.~\eqref{eq:stat1} reaches $\hat C_{\rm f}(\dot\gamma t,\dot\gamma t_w)$.
\section{Final Results for the FDR}\label{sec:FDres}
Our final result for the susceptibility in terms  of  the waiting time derivative reads,
\begin{equation}
\chi_{\rm f}(t)\approx-\frac{\partial}{\partial t} C_{\rm f}(t)+\frac{1}{2}\left.\frac{\partial}{\partial t_w}C_{\rm f}(t,t_w)\right|_{t_w=0}.
\label{eq:deltwai}
\end{equation} 
Eq.~\eqref{eq:deltwai} states the connection of two very different physical mechanisms: The violation of the equilibrium FDT and the waiting time dependence of the two-time correlator at $t_w=0$. The connection can be tested in simulations, where both quantities are accessible independently \cite{Berthier02,Zausch08}.  The extra term in the FDT can indeed be connected to the time derivative of a correlation function reflecting its dissipative character, but no such simple relation occurs as in equilibrium. Using our approximation in Eq.~\eqref{eq:lti} for the waiting time derivative, we can hence finally write our extended FDT
\begin{equation}
\chi_{\rm f}(t)\approx-\dot C_{\rm f}(t)+\frac{1}{2}\left(\dot C^{(t)}_{\rm f}(t)-\dot C^{(e)}_{\rm f}(t)\frac{C^{(t)}_{\rm f}(t)}{C^{(e)}_{\rm f}(t)}\right).\label{eq:FDT}
\end{equation}
This equation connects the susceptibility to measurable quantities, at least in simulations, without adjustable parameter.  \subsection{FDR as function of time}\label{sec:final}
\begin{figure}
\begin{center}\includegraphics[width=1\linewidth]{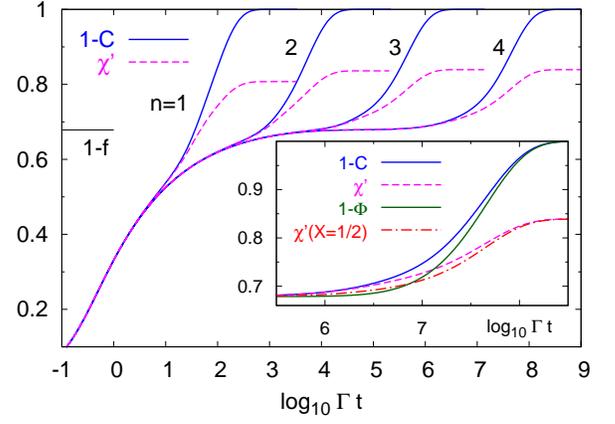}\end{center}
\caption{$C(t)$ from the $F_{12}^{(\dot\gamma)}$-model with Eq.~\eqref{eq:stat1} and $\chi(t)$ via Eq.~(\ref{eq:FDT}) for a glassy state ($\varepsilon=10^{-3}$) and $\dot\gamma/\Gamma=10^{-2n}$ with $n=1...4$. Shown are integrated correlation, $1-C(t)$ and response $\chi'(t)=\int_0^t\chi(t')dt'$. Inset shows additionally the transient correlator $\Phi$ for comparison and the $\hat X^{({\rm univ})}=\frac{1}{2}$ susceptibility for $\dot\gamma/\Gamma=10^{-8}$. From Ref.~\cite{Krueger09}.}
\label{fig:schem1}
\end{figure}
\begin{figure}[t]
\begin{center}\includegraphics[width=1\linewidth]{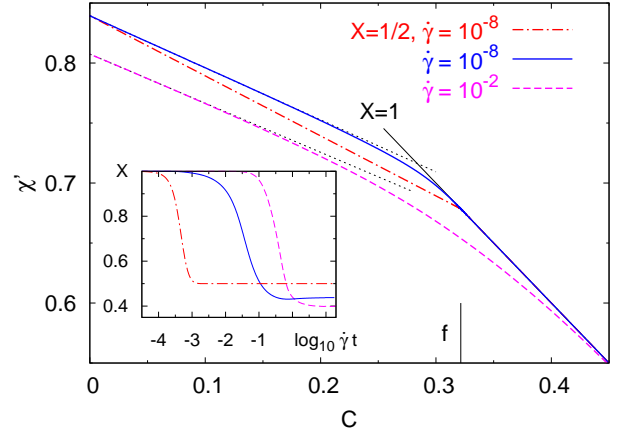}\end{center}
\caption{Parametric plot of correlation $C(t)$ versus response
$\chi'(t)=\int_0^t\chi(t')dt'$ from Fig.~\ref{fig:schem1} ($\varepsilon=10^{-3}$) 
together with constant non-trivial FDR (straight lines) at long times. The vertical solid line marks the plateau $f$. Inset shows the FDR $X(t)$ as function of
strain for the same susceptibilities. From Ref.~\cite{Krueger09}.}
\label{fig:schem2}
\end{figure}
\begin{figure}
\begin{center}\includegraphics[width=1\linewidth]{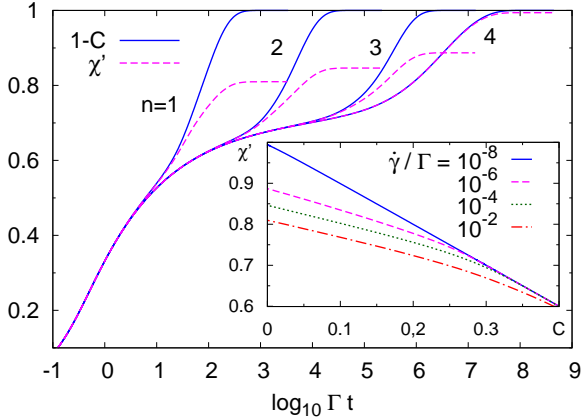}\end{center}
\caption{\label{fig:tefffl}$C(t)$ and $\chi(t)$ via Eq.~(\ref{eq:FDT}) for a fluid state ($\varepsilon=-10^{-3}$) and $\dot\gamma/\Gamma=10^{-2n}$ with $n=1...4$. Shown are integrated correlation, $1-C(t)$ and response $\chi'(t)=\int_0^t\chi(t')dt'$. Inset shows the parametric plot for the different shear rates.}
\end{figure}
The relation between susceptibility and correlators requires correlators as input.
We want to visualize the susceptibility using the schematic model \eqref{eq:f12}, Eq.~\eqref{eq:stat1} for the stationary correlator and Eq.~\eqref{eq:FDT}.
Fig.~\ref{fig:schem1} shows the resulting $\chi$ together with $C$ for a glassy state  at different shear
rates. For short times, the equilibrium FDT is valid, while for long
times the susceptibility is smaller than expected from the
equilibrium FDT. This deviation is qualitatively similar for the
different shear rates. For the smallest shear rate, we also plot
$\chi$ calculated by Eq.~(\ref{eq:FDT}) with
$\dot C_{\rm f}^{(t)}$ replaced by $\dot C_{\rm f}$, see inset of Fig.~\ref{fig:schem1}. In this approximation,  the FDR intriguingly takes the universal value $\hat X^{({\rm univ})}(\dot\gamma t)=\frac{1}{2}$, without any free parameters. The realistic susceptibility is achieved by including the difference between $C^{(t)}(t)$ and $C(t)$. The parameter $\tilde\sigma$ is directly proportional to this difference. 

In the parametric plot (Fig.~\ref{fig:schem2}), the $X=\frac{1}{2}$-approximation leads to two perfect lines with slopes $-1$ and $-\frac{1}{2}$ connected by a sharp kink at the
non ergodicity parameter $f$. For the realistic curves, this
kink is smoothed out, but the long time part
is still well described by a straight line, i.e., the FDR is still almost constant during the final relaxation process. We predict a non-trivial time-independent FDR $\hat X_{\rm f}(\dot\gamma t)=$const. if $C_{\rm f}^{(t)}$ (and with Eq.~(\ref{eq:stat1}) also $C_{\rm f}$) decays exponentially for long times, because $\Delta\chi_{\rm f}$ then decays exponentially with the same exponent. The slope of the long time line becomes smaller with increasing $\tilde\sigma$ (i.e., also with increasing value of $G_\infty/f$ in Eq.~\eqref{eq:mod}). We find that the value of the long time  FDR is always smaller than $\frac{1}{2}$ in glassy states. 

The line cuts the FDT line {\it below} $f$ for $\dot\gamma\to 0$. All these
findings are in excellent agreement with the data in Ref.~\cite{Berthier02}. The FDR itself is of interest also, as
function of time (inset of Fig.~\ref{fig:schem2}). A rather sharp transition from 1 to $\frac{1}{2}$ is observed when $\Phi\approx C$ is approximated,
which already takes place at $\dot\gamma t\approx10^{-3}$, a time
 when the FDT violation is still invisible in Fig.~\ref{fig:schem1}.
For the realistic curves, this transition happens two decades later. Strikingly, the huge difference is not apparent in the parametric
plot, which we consider a serious drawback of this representation.

Fig.~\ref{fig:tefffl} shows $C(t)$ and $\chi(t)$ for a fluid state. For large shear rates, these curves are similar to the glassy case, while for $\dot\gamma\to0$, the equilibrium FDT holds for all times. In the parametric plot (inset of Fig.~\ref{fig:tefffl}) one sees that the long time FDR is still approximately constant in time for the case $\dot\gamma\approx \tau^{-1}$ ($n=3$ in Fig.~\ref{fig:tefffl}), where shear relaxation and structural relaxation compete. $\tau$ is the $\alpha$ relaxation time of the un-sheared fluid.

Summarizing, we find that the two separated relaxation steps \cite{Fuchs03,Varnik06b} (Fig.~\ref{fig:twotimefull}) of the correlator in the limit of small shear rates for glassy states are connected to two different values of the FDR. During the shear
independent relaxation onto the plateau of height given by the
non-ergodicity parameter $f_{\rm f}$, we have $C^{(t)}_{\rm f}(t)= C^{(e)}_{\rm f}(t)+\mathcal{O}(\dot\gamma t)$ \cite{Fuchs03} in Eq.~\eqref{eq:FDT},
and the equilibrium FDT holds. During the shear-induced final
relaxation from $f_{\rm f}$ down to zero, i.e., for $\dot\gamma\to0$, and
$t\to\infty$ with $t \dot\gamma= $ const., the correlator without
shear stays on the plateau and its derivative is negligible. A
non-trivial FDR follows. In the glass holds
\begin{equation}\label{eq:shandlo}
\lim_{\dot\gamma\to 0} \chi_{\rm f}(t)=
\left\{ \begin{array}{ll}
-\dot C_{\rm f}(t) & \dot\gamma t\ll 1,\\
-\dot C_{\rm f}(t)+\frac{1}{2}\dot C^{(t)}_{\rm f}(t) & \dot\gamma t ={\cal O}(1),\\
=-\hat X_{\rm f}(\dot\gamma t) \frac{\partial}{\partial t}\hat C_{\rm f}(\dot\gamma t).
\end{array}\right.
\end{equation}
If one approximates stationary and
transient correlator to be equal \cite{Fuchs03}, $C^{(t)}_{\rm f}(t)\approx C_{\rm f}(t)$, we
find the interesting universal $X=\frac{1}{2}$-law for long times,
\begin{equation}
\lim_{\dot\gamma\to 0}\chi_{\rm f}(t\to\infty)=-\frac{1}{2}\dot C_{\rm f}(t). 
\end{equation}
The FDR, in this case, takes the universal value
$\lim_{\dot\gamma\to0}X_{\rm f}(t\to\infty)=\hat X^{({\rm univ})}(\dot \gamma
t)=\frac{1}{2}$, independent of $f$. This is in
good agreement with the findings in Ref.~\cite{Berthier02}, and corresponds to an effective temperature of $T_{\rm eff}/T=2$ for all observables. The initially
additive correction in Eq.~(\ref{eq:genFDT}) hence turns then into a
multiplicative one, which does not depend on rescaled time during
the complete final relaxation process. As summarized in Sec.~\ref{sec:FDTFDR}, many spin models yield $X=\frac{1}{2}$ at the critical temperature. The deviation from the value $\frac{1}{2}$ of the long time FDR in our approach comes from the difference between stationary and transient correlators.
\subsection{FDR as Function of Shear Rate}
\begin{figure}[t]
\begin{center}\includegraphics[width=1\linewidth]{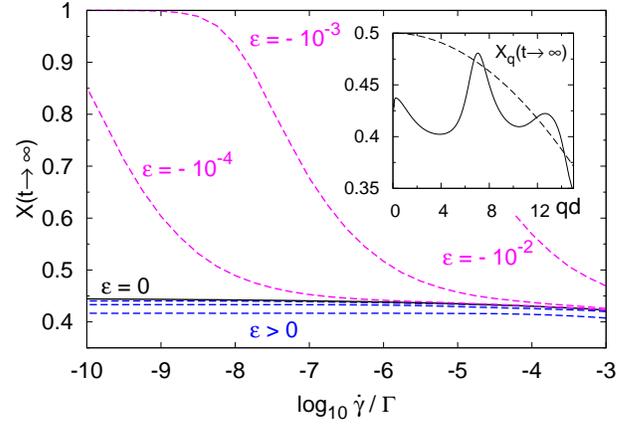}\end{center}
\caption{Long time FDR as function of shear rate for glassy
($\varepsilon=10^{-n}$) and liquid ($\varepsilon=-10^{-n}$) states with $n=2,3,4$. $X(t\to\infty)$ is determined from fits to the parametric plot as shown in Figs.~\ref{fig:schem1} and \ref{fig:tefffl}. Inset shows the long time FDR as function of wavevector $q$ for coherent (solid line) and incoherent (dashed line) density fluctuations at the critical density ($\varepsilon=0$).}
\label{fig:schem3}
\end{figure}
Fig.~\ref{fig:schem3} shows the long time FDR as a function of shear
rate for different densities above and below the glass transition. The FDR was 
determined via fits to the parametric plot in the interval $C(t)\in$ [0 :
0.1]. In the glass $X(t\to\infty)$ is nonanalytic while it goes to
unity in the fluid as $\dot\gamma\to0$ (compare Fig.~\ref{fig:tefffl}). We verified that the FDT-violation starts quadratic in $\dot\gamma$ in the fluid, as is to be expected due to symmetries. $X(t\to\infty)$ is also nonanalytic as function of $\varepsilon$ and jumps to an finite value less than one. For all densities, the FDR decreases with shear rate. For constant shear rate, it decreases with the density. This is also in agreement with the simulations \cite{Berthier02}.
\subsection{FDR as Function of Wavevector}
The realistic version of the extended FDT, taking into account the difference of transient and stationary correlator, gives an observable dependent FDR in general. This can be quantified by using the exponential approximation for the long time transient correlator for glassy states (compare Eq.~\eqref{eq:corr1}) \cite{Fuchs03}
\begin{equation}
C_{\rm f}^{(t)}(t\to\infty)\approx f_{\rm f}\,e^{-a_{\rm f}|\dot\gamma| t}. \label{eq:corr2}
\end{equation}
The long time FDR then follows with Eqs. \eqref{eq:FDT}, \eqref{eq:stat1} and \eqref{eq:lti}, 
\begin{equation}
X_{\rm f}(t\to\infty)=\frac{\frac{1}{2}-a_{\rm f}\tilde\sigma}{1-a_{\rm f}\tilde\sigma}.\label{eq:vonq}
\end{equation}
The inset of Fig.~\ref{fig:schem3} shows the long time FDR for coherent and incoherent density fluctuations at the critical density. We used the isotropic long time approximations \eqref{eq:corr1} and \eqref{eq:corrinc} for $a_{\rm f}$ respectively and $c\, \tilde\sigma=0.1$ from Eq.~\eqref{eq:mod}. The incoherent  case was most extensively studied in Ref.~\cite{Berthier02}. The FDR in Fig.~\ref{fig:schem3} is isotropic in the plane
perpendicular to the shear direction but not independent of wave vector $q$, contradicting the idea of an effective temperature as proposed in Refs.~\cite{Berthier99,Berthier02} and others. 

For $q\to\infty$, $h_q/f_q$ (corresponding to $a_{\rm f}$) grows without bound and the FDR in Eq.~\eqref{eq:vonq} becomes negative eventually. For the parameters we used, the root is at $q\approx 30$. For larger values of $c\tilde\sigma$ (i.e., for larger values of $G_\infty/f$ in \eqref{eq:mod}), the root is at smaller values of $q$.
According to our considerations in the discussion section and the available simulation data, a negative FDR is unphysical. For large values of $q$, the exponential approximation for the transient correlator or our approximation \eqref{eq:stat1} for the two-time correlator  might not be justified.
\subsection{Direct Comparison to Simulation Data}
\begin{figure}[t]
\begin{center}\includegraphics[width=1\linewidth]{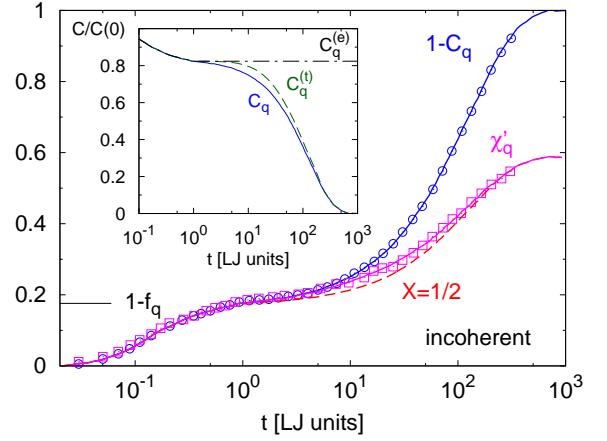}\end{center}
\caption{\label{fig:fit}Comparison to simulation data for incoherent
density fluctuations in the neutral direction (wave vector ${\bf q}=7.47 {\bf
e}_z$) at temperature $T=0.3$ ($T_c=0.435$) and $\dot\gamma=10^{-3}$. Circles and squares are the data
(including units) from Fig.~11 in Ref.~\cite{Berthier02}, lines are
$1-C_{{\bf q}}$ from Fig.~8 in Ref.~\cite{Berthier02}, and the response
$\chi_{\bf q}'(t)=\int_0^t\chi_{\bf q}(t')dt'$ calculated via Eq.~(\ref{eq:FDT}).  The dashed line shows $\chi_{\bf q}'$ with
approximation $C^{(t)}_{\bf q}\approx C_{\bf q}$. Inset shows the different correlators, see
main text.}
\end{figure}
Despite the dependence of the long time FDR on wavevector, Eq.~(\ref{eq:FDT}) is not in contradiction to the data in Ref.~\cite{Berthier02}, as can be seen by direct comparison to
their Fig.~11. For this, we need the quiescent as well as the
transient correlator as input. $C^{(e)}_{q}$ has been measured in
Ref.~\cite{Varnik06b}, suggesting that it can be approximated by a
straight line beginning on the plateau $f_q$ of $C_{{\bf q}}(t)$. In Fig.~\ref{fig:fit} we show the resulting susceptibilities. There is no
adjustable parameter, when $C^{(t)}_{\bf q}\approx C_{\bf q}$ is
taken. For the other curve, we calculated $C^{(t)}_{{\bf q}}(t)$ 
by inversion of Eq.~(\ref{eq:stat1}). We used the dimensionless number $\tilde\sigma=0.01$ as fit parameter which was chosen such that the resulting susceptibility fits best with the simulation data. The achieved agreement to $\chi$ from the simulations is striking. In the inset we show the original $C_{{\bf q}}$ from Ref.~\cite{Berthier02} together with our construction of $C^{(e)}_{q}$
and the calculated $C^{(t)}_{{\bf q}}$. It which appears very
reasonable compared with recent simulation data on $C_{\rm f}(t,t_w)$ \cite{Zausch08} and compared with Fig.~\ref{fig:twotimefull}. The value for $f_q$ used to construct $C^{(e)}_{q}$ is also indicated in the main figure.
\section{Hermitian Part of the Smoluchowski Operator and Comoving Frame}\label{sec:Corr}
In this section, we want to understand the violation of the FDT from a different point of view, i.e., from exact reformulations of the starting point \eqref{eq:sus}. First, we split the SO into Hermitian and anti-Hermitian part to see how the Hermitian part is connected to the susceptibility in Eq.~\eqref{eq:sus}. We will then see that one can reformulate \eqref{eq:sus} in terms  of an advected derivative.
\subsection{Hermitian part}
Investigating the stationary correlator in Eq.~\eqref{eq:stationary}, one finds that  the operator $\Omega^\dagger$ is not Hermitian in the  average with $\Psi_s$ \cite{graham}. This is why one cannot show that $C_{\rm f}(t)$ is of positive type \cite{McLennan} via, e.g. an expansion in eigenfunctions since only an expansion into a biorthogonal set is possible \cite{Risken, Fuchs05}.

Subsequent to realizing this, we want to split the SO into its Hermitian and its anti--Hermitian part with respect to the average with $\Psi_s$. Recall that $\Omega$ is the adjoint of $\Omega^\dagger$ in the unweighted scalar product \cite{Dhont, Risken, Fuchs05}. The adjoint of $\Omega^\dagger$ in the stationary average is defined by
\begin{eqnarray}
\left\langle g \,\Omega^\dagger\, f^*\right\rangle^{(\dot\gamma)}=\int d\Gamma \Psi_s(\Gamma) g(\Gamma)\,\Omega^\dagger f^*(\Gamma)\nonumber\\=\int d\Gamma \Psi_s(\Gamma) f^*(\Gamma)\,\tilde\Omega^\dagger g(\Gamma)=\left\langle f^* \,\tilde\Omega^\dagger \,g\right\rangle^{(\dot\gamma)}.
\end{eqnarray}
We already stressed that $\tilde\Omega^\dagger$ is neither identical to $\Omega$ nor $\Omega^\dagger$, 
\begin{equation}
\tilde\Omega^\dagger=\sum_i \boldsymbol{\partial}_i^2+2\,(\boldsymbol{\partial}_i \ln \Psi_s)\cdot \boldsymbol{\partial}_i-{\bf F}_i\cdot \boldsymbol{\partial}_i-\boldsymbol{\partial}_i\cdot{\kap}\cdot{\bf r}_i.
\end{equation}
The difference between non-equilibrium forces $\boldsymbol{\partial}_i \ln \Psi_s$ and the potential forces $\vct{F}_i=\boldsymbol{\partial}_i \ln \Psi_e$ appears \cite{Szamel}.
Now the Hermitian and the anti-Hermitian parts of $\Omega^\dagger$ with respect to stationary averaging are given by
\begin{subequations}
\begin{align}
\Omega^\dagger_{H}=&\frac{1}{2}(\Omega^\dagger+\tilde\Omega^\dagger)= \sum_i \boldsymbol{\partial}_i^2+(\boldsymbol{\partial}_i \ln \Psi_s)\cdot\boldsymbol{\partial}_i\,,\\
\Omega^\dagger_{A}=&\frac{1}{2}(\Omega^\dagger-\tilde\Omega^\dagger)\nonumber\\=&\sum_i-(\boldsymbol{\partial}_i \ln \Psi_s)\cdot\boldsymbol{\partial}_i+{\bf F}_i\cdot \boldsymbol{\partial}_i+\boldsymbol{\partial}_i\cdot{\kap}\cdot{\bf r}_i\,.\label{eq:omegaa}
\end{align}
\end{subequations}
We obviously have $\Omega^\dagger=\Omega^\dagger_{H}+\Omega^\dagger_{A}$.
$\Omega^\dagger_{H}$ is similar to the equilibrium SO $\Omega^\dagger_e$ with forces $\vct{F}_i$ replaced by the non-equilibrium forces  $\boldsymbol{\partial}_i \ln \Psi_s$. As expected, the anti-Hermitian part contains the shear part $\delta\Omega^\dagger$. It also contains the difference between equilibrium and non-equilibrium forces.
The eigenvalues of $\Omega^\dagger_{A}$ are imaginary and the eigenvalues of $\Omega_H^\dagger$ are real \cite{Risken}. In the given case, $\Omega^\dagger_H$ can furthermore be shown to have negative semi-definite spectrum as does the equilibrium operator, because we have
\begin{equation}
\left\langle f^* \Omega^\dagger_{H} g\right\rangle^{(\dot\gamma)}=-\left\langle \frac{\partial f^*}{\partial {\bf r}_i}\cdot \frac{\partial g}{\partial {\bf r}_i}\right\rangle^{(\dot\gamma)}.\label{eq:negeig}
\end{equation}
If the correlation function is real for all times, $C_{\rm f}(t)=C^*_{\rm f}(t)$, as can be shown e.g. for density fluctuations \cite{Fuchs09}, the initial decay rate is negative since $\Omega^\dagger_A$ does not contribute,
\begin{equation}
\mathcal{R} \left\{\left\langle f^* \Omega^\dagger f\right\rangle^{(\dot\gamma)}\right\}=\left\langle f^* \Omega^\dagger_{H} f\right\rangle^{(\dot\gamma)}=-\left\langle \frac{\partial f^*}{\partial {\bf r}_i} \cdot\frac{\partial f}{\partial {\bf r}_i}\right\rangle^{(\dot\gamma)}.\label{eq:inicomov}
\end{equation}
Thus a real correlator initially always decays, i.e., the external shear cannot enhance the fluctuations. Higher order terms in $t$ contain contributions of $\Omega^\dagger_A$ and such an argument is not possible. 
\subsection{Susceptibility and comoving frame}
We now come to the connection of $\Omega^\dagger_H$ to the susceptibility. Eq.~\eqref{eq:sus} can be written
\begin{equation}
\chi_{\rm fg}=-\left\langle f^* \Omega^\dagger_{H} \,e^{\Omega^\dagger t} g\right\rangle^{(\dot\gamma)}.\label{eq:chiherm}
\end{equation}
The response of the system is not given by the time derivative with respect to the full dynamics but by the time derivative with respect to the Hermitian, i.e., the ``well behaved'' dynamics. 
It follows that we can write the susceptibility,
\begin{equation}
\chi_{\rm fg}=-\left\langle f^* (\Omega^\dagger-\Omega^\dagger_A)\,e^{\Omega^\dagger t} g\right\rangle^{(\dot\gamma)}.
\end{equation}
We note that this equation is very similar to Eq.~(13) in Ref.~\cite{Baiesi09} since $\frac{1}{2}(L-L^*)$ is the anti-Hermitian part of $L$ in Ref.~\cite{Baiesi09}. Eq.~\eqref{eq:chiherm} can be made more illustrative by realizing that $\Omega^\dagger_A$ can be expressed by the probability current ${\bf j}_i^s$,
\begin{equation}
\Omega^\dagger_A=\Psi_s^{-1} \sum_i\vct{j}_i^s \cdot\boldsymbol{\partial}_i.
\end{equation} 
We hence finally have
\begin{equation}
\chi_{\rm fg}=-\left\langle f^* \left(\partial_t- \Psi_s^{-1} \sum_i\vct{j}_i^s \cdot\boldsymbol{\partial}_i\right)\,e^{\Omega^\dagger t} g\right\rangle^{(\dot\gamma)}.\label{eq:comov}
\end{equation}
The derivative in the brackets can be identified as the convective or comoving derivative which is often used in fluid dynamics \cite{Landauh}. It measures the change of the function in the frame comoving with the probability current. If one could measure the fluctuations in this comoving frame, these would be connected to the corresponding susceptibility by the equilibrium FDT. This was also found for the velocity fluctuations of a single driven particle in Ref.~\cite{Speck06}. The difference in our system is that the probability current, i.e., the local mean velocity, speaking with the authors of Ref.~\cite{Speck06}, does not depend on spatial position $x$, but on the relative position of all the particles because it originates from particle interactions. 

Let us finish with interpreting the comoving frame. $X_{\rm f}$ describes the tendency of particles to move with the stationary current. If the stationary current vanishes, we have $X_{\rm f}=1$. If the particle trajectories are completely constraint to follow the current, we have $X_{\rm f}=0$, because a small external force cannot change these trajectories and $\chi_{\rm f}=0$. As examples for the latter case, let us speculate about the experiments in Refs.~\cite{Pine05,Gollub06}. They consider a rather dilute suspension of colloids in a highly viscous solvent. The bare diffusion coefficient is approximately zero (so called non-Brownian particles), i.e., on the experimental timescale the particles do not move at all without shear. Under shear, the particles move with the flow and one observes diffusion in the directions perpendicular to the shearing due to interactions. A very small external force does not change the trajectories of the particles (on the timescale of the experiment) due to the high viscosity. We expect $X_{\rm f}=0$ in this case because the particles completely follow the probability current. The studies in Ref.~\cite{Pine05,Gollub06} do not consider the susceptibility, the focus is put on the question whether the system is chaotic or not. The finding that the dynamics is irreversible under some conditions makes it even harder to predict the FDR, which would be of great interest.
\subsection{FDT for eigenfunctions}
From Eq.~\eqref{eq:chiherm} and $\Omega^\dagger_{H}=\frac{1}{2}(\Omega^\dagger+\tilde\Omega^\dagger)$, we find for arbitrary $f=f(\{x_i,y_i,z_i\})$,
\begin{align}
\chi_{\rm f} (t)=&-\frac{1}{2}\left\langle f^* \Omega^\dagger e^{\Omega^\dagger t}f\right\rangle^{(\dot\gamma)}-\frac{1}{2}\left\langle (\Omega^\dagger f^*)  e^{\Omega^\dagger t}f\right\rangle^{(\dot\gamma)},\label{eq:FDTnew}\\
=&-\frac{1}{2}\frac{\partial}{\partial t}\left\langle f^* e^{\Omega^\dagger t}f\right\rangle^{(\dot\gamma)}-\frac{1}{2}\left\langle (\Omega^\dagger f^*)  e^{\Omega^\dagger t}f\right\rangle^{(\dot\gamma)}.\nonumber
\end{align}
This form is especially illustrative since it explicitely shows that the FDT violation occurs because $\Omega^\dagger$ is not Hermitian in the stationary average. If it was, the two terms above would be equal and the equilibrium FDT would hold. We note that this form is equivalent to Eq.~(11) in Ref.~\cite{Baiesi09}. As pointed out by Baiesi, Maes and Wynants, Eq.~\eqref{eq:FDTnew} in the case of simulations has the advantage that correlation functions of well defined quantities ($f$ and $\Omega^\dagger f$) can be evaluated. This indicates  the usefulness of Eq.~\eqref{eq:FDTnew} relative to Eqs.~\eqref{eq:genFDT} and \eqref{eq:comov}. 

If we consider the case that $f=\phi_n$ with $\phi_n$ eigenfunction of $\Omega^\dagger$, $\Omega^\dagger\phi_n=\lambda_n\phi_n$, we find 
\begin{equation}
\chi_{\phi_n} (t)= -\frac{\partial}{\partial t} C_{\phi_n}(t).
\end{equation}  	
The equilibrium FDT thus holds for $f=\phi_n$.
\section{Discussion}\label{sec:disc}
\subsection{Deterministic versus Stochastic Motion}\label{sec:det}
We saw in Sec.~\ref{sec:Corr} that the susceptibility measures the fluctuations of the particles in the frame comoving with the probability current $\vct{j}_i^s$. We conclude that we  can split the displacements of the particles into two meaningful parts. First, the stochastic motion in the frame comoving with the average probability current. Second, the motion following the average probability current, which is deterministic and comes from the particle interactions.  The deterministic part is not measured by the susceptibility, $\chi$ is thus smaller than expected from the equilibrium FDT. It measures only parts of the dynamics. We have $X_{\rm f}\leq 1$.  
Let us quantify the above discussion as far as possible. In Eq.~\eqref{eq:chiherm}, we see that we can formally split the time derivative of the stationary correlation function into two pieces, the  stochastic one, measured by the susceptibility, and the deterministic one following the probability current,
\begin{equation}
\frac{\partial}{\partial t} C_{\rm f} (t)=\left\langle f^* \Omega^\dagger_H\,e^{\Omega^\dagger t} g\right\rangle^{(\dot\gamma)}+\left\langle f^* \Omega^\dagger_A\,e^{\Omega^\dagger t} g\right\rangle^{(\dot\gamma)}\nonumber
\end{equation}
The inequality $X_{\rm f}\leq \frac{1}{2}$ in glasses, see Fig.~\ref{fig:schem3} translates into an inequality for the two derivatives above,
\begin{equation}
\left|\left\langle f^* \Omega^\dagger_A\,e^{\Omega^\dagger t} g\right\rangle^{(\dot\gamma)}\right|\geq \left|\left\langle f^* \Omega^\dagger_H\,e^{\Omega^\dagger t} g\right\rangle^{(\dot\gamma)}\right|.
\end{equation}
In completely shear governed decay of glassy states, the deterministic displacements of the particles due to the probability current are larger than the stochastic fluctuations around this average current. In other words, if the stochastic motion was faster than the deterministic one, the decay would not be completely shear governed. 

It is likely that with increasing density or lowering temperature, the particles are more and more confined to follow the probability current and the FDR gets smaller and smaller and might eventually reach zero.
\subsection{Shear Step Model}
Trap models have repeatedly been used to study the slow dynamics of glassy systems and to investigate the violation of the equilibrium FDT \cite{Bouchaud92,Barrat95,Monthus96,Rinn00,Fielding02,Sollich02,Fielding00}. They boldly simplify the dynamics of super-cooled liquids and glasses, because the particles themselves form the traps for each other and it is therefore not easily possible to map the problem onto a single particle problem. Nevertheless, we want to introduce a simple toy model which will provide more insight into the FDT violation of sheared colloidal glasses. The model is depicted in 
\begin{figure}
\begin{center}
\includegraphics[width=0.6\linewidth]{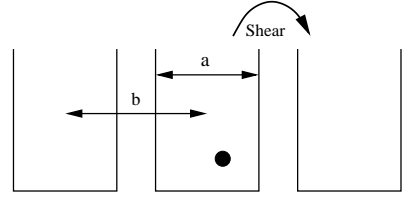}
\caption{\label{fig:toy}The shear step model: The particle is trapped in wells of width $a$ with infinitely high potential barriers. The center-to-center distance of the wells is denoted $b$.}
\end{center}
\end{figure}
Fig.~\ref{fig:toy}: The particle, surrounded by solvent (diffusivity $D_0=k_BT \mu_0$, we restore physical units), is trapped in an infinite potential well of width $a$. This potential well is one in a row of infinitely many with center-to-center distances $b$. The particle diffuses in the well and a very long time after we measured its position, its probability distribution $P(z)$ is constant within the well. 
The shear, i.e., a mechanism which lifts the particle over the potential barriers is introduced as follows; At time $t=\Theta, 2\Theta, 3\Theta, \dots$, the 'shear steps' lift the particle into a neighboring well according to its current position: If it is on the left side of the well, it gets to the well on the left hand side, if it is on the right hand side, it gets into the well on the right hand side. The model hence describes a direction perpendicular to the shear flow, e.g. the $z$-direction, where the influence of shear is symmetric. The initial position of the particle in the new well shall be distributed randomly. The resulting dynamics has some similarities to a Cauchy process \cite{Gardiner} and shares many properties of colloidal suspensions as will be shown below.  

Detailed balance is broken because the reverse step, that a particle is taken out of the right side of a trap and put back into its left neighbor, is missing.

The shear step model is much simpler than other trap models considered in the literature. The Bouchaud trap model \cite{Monthus96} contains a distribution of traps of different depth, allowing to study different situations such as aging. The simplicity of our model makes it easier to be analyzed and the result for the FDR contains only the parameters $a$ and $b$, whose values are of comparable size.

We regard the limit of small shear rates, it corresponds to $\Theta\gg a^2/D_0$, i.e., the time between two shear steps is much longer than it takes the particle to relax in the well. We will first present the mean squared displacement of the particle and then its mobility under a small test force.

At short times, the mean squared displacement (MSD) of the particle is the one of a free particle,
\begin{equation}
\lim_{t\to0}\langle (z-z(0))^2\rangle=2D_0t.\label{eq:dynsh}
\end{equation}
For times $\Theta>t\gg a^2/D_0$, the dynamics of the particle is glassy, i.e., the MSD is constant on the plateau. This plateau value can be derived from the constant probability distribution of the particle in the well, $P(z)=1/a$,
\begin{equation}
\langle (z-z(0))^2\rangle=\frac{1}{a^2}\int_{-a/2}^{a/2} dz(0)\int_{-a/2}^{a/2} dz (z-z(0))^2=\frac{a^2}{6}.\label{eq:inicond}
\end{equation}
We notice that in accordance with the glassy dynamics of colloidal suspensions, the plateau value is independent of $D_0$ and temperature. Note that the initial positions are distributed with $P(z)$ as well.
At long times, the particle performs a random walk with step-length $b$ and number of steps $t/\Theta$ \cite{footnote3} and the MSD approaches
\begin{eqnarray}
\langle (z-z(0))^2\rangle= b^2 \frac{t}{\Theta},\hspace{1cm}t\gg\Theta.\label{eq:longtimemsd}
\end{eqnarray}
The long time dynamics is independent of temperature and $D_0$, as is the long time decay of the density correlator for sheared colloidal glasses \cite{Fuchs03,Fuchs09}. The timescale is set by $\Theta$, corresponding to the a 'shear rate' of $\dot\gamma=\Theta^{-1}$. 
\begin{figure}
\begin{center}
\includegraphics[width=1\linewidth]{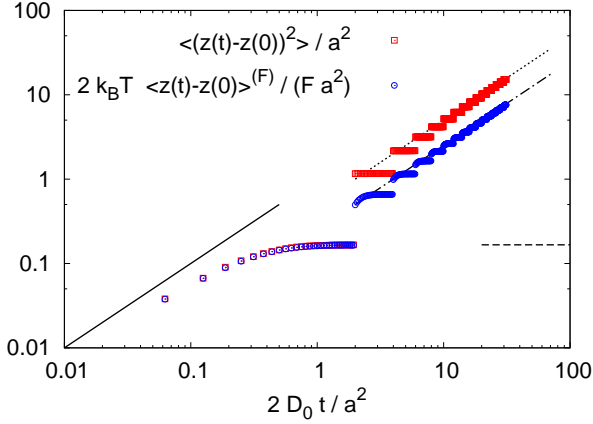}
\caption{\label{fig:simu1}
Mean squared displacement without (red squares) and mean traveled distance with external force $F$ (blue circles) in the shear step model with $b=a$ and $\Theta=2 a^2/(2D_0)$. Solid line shows the short time asymptote of Eqs.~\eqref{eq:dynsh} and \eqref{eq:mobsh}. Dashed line shows the plateau value of $\frac{a^2}{6}$, Eqs.~\eqref{eq:inicond} and \eqref{eq:mobint}. The dotted line shows the long time asymptote from Eq.~\eqref{eq:longtimemsd}, dash-dotted line the asymptote from Eq.~\eqref{eq:longtimemob}.  }
\end{center}
\end{figure}

To calculate the fluctuation dissipation ratio $X$, we have to find the mobility $\mu(t)$ in response to a small test force $F$ acting on the particle, say to the right. This test force shall not influence the jump rules defined above. At short times the particle obeys the (integrated) equilibrium FDT (the Einstein relation),
\begin{equation}
\lim_{t\to0} \frac{1}{F}\left\langle (z-z(0))\right\rangle^{(F)}=\lim_{t\to0}\frac{1}{2k_BT} \langle (z-z(
0))^2\rangle =\mu_0t.\label{eq:mobsh}
\end{equation} 
$\left\langle \dots\right\rangle^{(F)}$ denotes an average under the influence of
 the external force. The external force changes the probability distribution of the particle in the well. It is more likely to find the particle on the right hand side of the well than on the left hand side. The distribution $P_F(z)$ for $\Theta>t\gg a^2/D_0$ follows the Boltzmann distribution \cite{Dhont}, which, in linear order of the external force, reads,
\begin{equation}
P_F(z)=\left\{\begin{array}{cc}
\frac{1}{a}\left(1+\frac{F\,z}{k_BT}\right)&-a/2<z<a/2,\\
0&\rm else.
\end{array}\right.\hspace{0cm} 
\label{eq:probdist}
\end{equation}
The mean traveled distance is easily derived,
\begin{align}
&\frac{1}{F}\left\langle (z-z(0))\right\rangle^{(F)}=\\&\frac{1}{a}\int_{-a/2}^{a/2} dz(0) \int_{-a/2}^{a/2} dz P_F(z) (z-z(0))=\frac{a^2}{12 k_BT}.\label{eq:mobint}
\end{align}
Comparing with Eq.~\eqref{eq:inicond}, the equilibrium FDT holds for all times $t<\Theta$ as expected.

Due to the distorted probability distribution $P_F(z)$ in Eq.~\eqref{eq:probdist}, the shear step at $t=\Theta$ will take the particle more likely to the right.  The rate $R$ for jumps to the right minus the rate for jumps to the left follows,
\begin{equation}
R=\frac{\int_{0}^{a/2} P_F(z) dz- \int_{-a/2}^{0} P_F(z) dz}{\int_{-a/2}^{a/2} P_F(z) dz}=\frac{a}{4}\frac{F}{k_BT}+ \mathcal{O}(F^2).
\end{equation}
At every shear step, the particle travels on average the distance $b R$. For $t\gg\Theta$, the initially traveled distance in the well is negligible and we have
\begin{equation}
\frac{1}{F}\left\langle (z-z(0))\right\rangle^{(F)}= \frac{a\,b}{4 k_BT}\frac{t}{\Theta}, \hspace{1cm}t\gg\Theta.\label{eq:longtimemob}
\end{equation}
The mobility of the particle is finite at long times and independent of the diffusivity $D_0$. The FDR $X$ is in this situation defined by
\begin{equation}
\frac{1}{F}\frac{\partial}{\partial t}\left\langle (z-z(0))\right\rangle^{(F)}=X(t)\frac{1}{2k_BT} \frac{\partial}{\partial t}\langle (z-z(0))^2\rangle.
\end{equation}
We illustrate the results above with simulations of the described dynamics, see Appendix \ref{sec:detsimu} for details. Fig.~\ref{fig:simu1} shows the simulation results together with the derived asymptotic formulae. We see that the FDT holds for times $t<\Theta$ and is violated for $t>\Theta$.
\subsubsection{Long Time FDR}
\begin{figure}
\begin{center}                                                                       \includegraphics[width=1\linewidth]{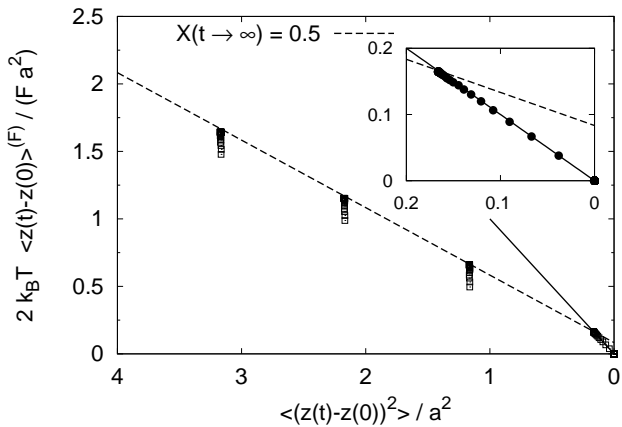}
\caption{\label{fig:simu2}
Parametric plot of MSD versus mean traveled distance for $b=a$ and $\Theta=2 a^2/(2D_0)$. Solid line shows the short time relation $X=1$. Dashed line shows the long time FDR $X=\frac{1}{2}$. Inset shows the same graph zoomed into the short time part.}
\end{center}
\end{figure}

We see in Fig.~\ref{fig:simu1} that the Einstein relation does not hold for long times. The long time FDR $X(t\gg\Theta)$ is different from unity and follows with Eqs.~\eqref{eq:longtimemsd} and \eqref{eq:longtimemob},
\begin{equation}
X(t)=\left\{\begin{array}{cc}
1&t<\Theta,\\
\frac{1}{2}\frac{a}{b}& t\gg\Theta,
\end{array}\hspace{1cm}\Theta\gg\frac{a^2}{D_0}.\right.\label{eq:longtimeFDR}
\end{equation}
This is illustrated in Fig.~\ref{fig:simu2}.
\subsubsection{Discussion}
Let us emphasize the similarities between the shear step model and the sheared colloidal glass. For both, $X(t\gg\Theta)$ is independent of temperature because the potentials are infinitely high. In soft sphere glasses, this is not true because the temperature also governs the glass transition. 

In the colloidal glass, the particle is trapped in the cage. The short time motion, the rattling in the cage, is FDT-like, as is the motion of the particle in the potential well. For long times, the shear drives the particle into a neighboring cage. For the directions perpendicular to the shear, this motion is symmetric (without test force) as in the shear step model. In the shear step model, the small external force does not change the shear step mechanism, it only changes the probability distribution of the particle in the well and thereby the motion becomes asymmetric. It is appealing to imagine something similar to happen in the colloidal system: The external force influences the distribution in the cage and makes the probability for the particle to be driven to the neighboring cages asymmetric. In both cases, this mechanism leads to a finite mobility which does not obey the equilibrium FDT.
 
In the shear step model, $X(t\gg\Theta)$ reaches $\frac{1}{2}$ if $b$ goes to $a$. $b<a$ (leading to $1>X>\frac{1}{2}$) is physically not reasonable and we get the constraint of $X\leq\frac{1}{2}$ which coincides with the one for the real system, see Fig.~\ref{fig:schem3}.

The model allows to discuss two more effects: In the colloidal system at increasing density, the cages become smaller and smaller. This corresponds to a decreasing value of $\frac{a}{b}$, and $X(t\gg\Theta)$ decreases. This is in accordance with Fig.~\ref{fig:schem3} and Ref.~\cite{Berthier02}. In the shear step model, it eventually reaches zero being always positive. 
\begin{figure}
\begin{center}
\includegraphics[width=1\linewidth]{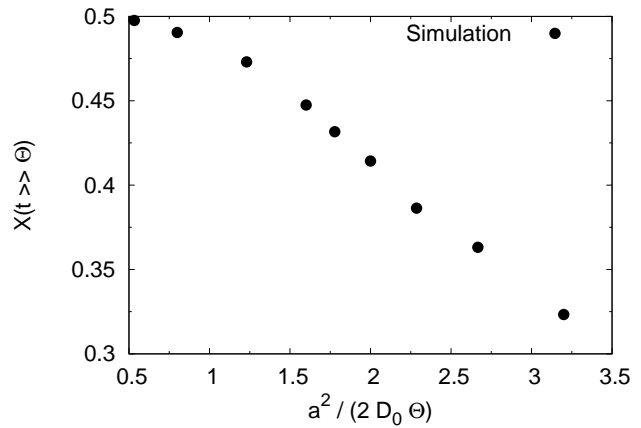}
\caption{\label{fig:simu4}
Long time FDR in the shear step model as function of 'shear rate' $1/\Theta$.}
\end{center}
\end{figure}

For increasing shear rates, i.e., decreasing $\Theta$, the particle has less time to relax in the well and the effect of the external force decreases. Fig.~\ref{fig:simu4} shows that decreasing $\Theta$ lowers the value of $X$. It is also in agreement with Fig.~\ref{fig:schem3} and the simulations in Ref. \cite{Berthier02}. We believe that the decrease in the colloidal system has the same origin, i.e., the particle has less time to adjust its distribution in the cage in response to the force. 

The FDT has been studied in Bouchaud's trap model previously \cite{Fielding02}, where in contrast to our model, the parametric plot has continously varying slope. The model in Ref.~\cite{Fielding02} is more realistic than ours and different observables can be studied. The advantage of the shear step model is its simplicity and its time independent long time FDR which has only the physically illustrative free parameters $a$ and $b$.

\section{Summary}\label{sec:summary}
We investigated the relation between susceptibility and correlation functions for colloidal suspensions at the glass transition. While the equilibrium FDT holds at short times, a time-independent positive FDR smaller than unity is obtained at long times during the shear driven decay. We find that the long time FDR is nearly isotropic in the plane perpendicular to the shear flow and takes the universal
value $\hat X(\dot\gamma t)=\frac{1}{2}$ in glasses
at small shear rates in the simplest approximation. This agrees with the interpretation of an effective temperature. Nevertheless, corrections
arise from the difference of the stationary to the transient
correlator and depend on the considered observable. They alter
$\hat X_{\rm f}$ to values $\hat X_{\rm f}\leq\frac{1}{2}$ in the
glass. Our findings are in good agreement with the simulations in Ref.~\cite{Berthier02}. 

While we used as central approximation the novel relation for the waiting time derivative, Eq.~\eqref{eq:lti}, a more standard MCT and projection operator approach leads to qualitatively equivalent results; see Appendix \ref{sec:ZM}. Considering the crudeness of the MCT decoupling and of our approximation \eqref{eq:lti}, the quantitative differences between the two approaches appear reasonable. From both approaches we can conclude that there is a nontrivial FDR $\hat X_{\rm f}(\dot\gamma t)=\hat X_{\rm f}$ for small shear rates during the final shear driven decay.

$\hat X_{\rm f}$ depends on shear rate non-analytically for all shear-melted glassy states. At the glass transition, the $\dot\gamma\to 0$ value jumps discontinuously from its nontrivial value $\hat X_{\rm f}(\dot\gamma\to0) < 1$ to the equilibrium value $X_{\rm f}=1$. For finite drive, $\hat X_{\rm f}$ decreases below unity for all states. The discontinuous behavior of $\hat X_{\rm f}(\dot\gamma\to 0)$ results from the shear driven decay on timescale $\dot\gamma^{-1}$. Within MCT-ITT the shear governed final decay is also the origin of a finite dynamic yield stress which also jumps discontinuously to its equilibrium value at the glass transition. These predictions differ from the mean-field spin-glass results. Ref.~\cite{Berthier02} finds power-law-fluid behavior but no dynamic yield stress. Moreover, the FDR at vanishing shear rate moves continously to the equilibrium value unity at the glass transition. The MCT-ITT scenario for yielding and fluctuation dissipation relations is thus unique compared to other approaches to shear-melted glasses. Investigation of shear driven glassy dispersions thus provides a unique possibility to discriminate between different theories of the glass transition.

The incoherent motion and the relation between diffusivity and mobility can also be studied within our approach, and is topic of a companion paper \cite{Krueger09d}.

Our finding of values close to the universal $\hat
X(\dot\gamma t)=\frac{1}{2}$ points to intriguing connections to critical spin models. Open questions concern establishing such a connection and to address the
concept of an effective temperature, which
was developed for ageing and driven mean field models. It might also be interesting in the future to study the response of the system to small perturbations in the shear rate $\dot\gamma$ \cite{Speck09}.

We derived a relation between the dominant part of the violating term and the waiting time derivative of $C_{\rm f}(t,t_w)$, viz a relation of two completely different physical quantities. This connection can be tested in simulations. Because we identified all but one contributions of the violating term with independently measurable quantities, all our approximations are testable independently in simulations. The difference of the measured $\Delta\chi_{\rm f}$ in FDT simulations to the sum of the measured terms of $\Delta\chi_{\rm f}$ in waiting time simulations yields the contribution of the last term in $\Delta\chi_{\rm f}$.

We presented a new exact formulation of the susceptibility which involves the Hermitian part of the SO. This part is interpreted to represent the dynamics in the frame comoving with the probability current. 
 
We introduced the shear step model to illustrate the FDT violation in sheared colloidal suspension. In the shear step model, the FDR takes values $X\leq\frac{1}{2}$. 

\begin{acknowledgments}
We thank A.~Gambassi, P.~Sollich, M.~E.~Cates and G.~Szamel for helpful discussions.
M.~K. was supported by the Deutsche Forschungsgemeinschaft in the International Research and Training Group 667 ``Soft Condensed Matter of Model Systems''. 
\end{acknowledgments}

\begin{appendix}
\section{Mode Coupling Approach}\label{sec:ZM}
In this Appendix, we start over and present the analysis of the susceptibility in Eq.~\eqref{eq:deltachi} for density fluctuations using Zwanzig Mori projections. $f=\varrho_{\bf q}=\sum_i e^{i{\bf q}\cdot{\bf r}_i}$ for  coherent and $f=\varrho^s_{\bf q}=e^{i{\bf q}\cdot{\bf r}_s}$ for incoherent density fluctuations. We will treat the two cases at once with $\varrho$ denoting either $\varrho_{\bf q}$ or $\varrho^s_{\bf q}$. $f$ being $x$-independent translates into $q_x=0$. Normalized equilibrium, transient and stationary correlators, as defined in Sec.~\ref{sec:correlation}, are denoted $\Phi_q^{(e)}(t)=\langle \varrho^* e^{\Omega^\dagger_e t} \varrho\rangle/\langle\varrho^*\varrho\rangle$, $\Phi_{\bf q}(t)=\langle \varrho^* e^{\Omega^\dagger t} \varrho\rangle/\langle\varrho^*\varrho\rangle$ and $C_{\bf q}(t)=\langle \varrho^* e^{\Omega^\dagger t} \varrho\rangle^{(\dot\gamma)}/\langle\varrho^*\varrho\rangle^{(\dot\gamma)}$. Stationary averages are normalized with $\langle \varrho^*\varrho\rangle^{(\dot\gamma)}$, the initial value of the stationary correlator. In the coherent case, this is the distorted static structure factor $S^{(\dot\gamma)}_{\bf{q}}=\langle \varrho_{\bf q}^* \varrho_{\bf q}\rangle^{(\dot\gamma)}/N$, in the incoherent case, it equals unity, $\varrho_{\bf q}^{s*}\varrho_{\bf q}^s=1$. The transient coherent correlator is normalized by the equilibrium structure factor $S_q=\langle \varrho_{q}^* \varrho_{q}\rangle/N$. The normalized violating term is then defined by (compare Eq.~\eqref{eq:sus})
\begin{equation}
\Delta\chi_{\bf q}(t)=\frac{1}{\langle \varrho^*\varrho\rangle^{(\dot\gamma)}}\left\langle \sum_i \frac{\partial  \varrho^*}{\partial {\bf r}_i}\cdot\boldsymbol{\partial}_i e^{\Omega^\dagger t} \varrho\right\rangle^{(\dot\gamma)}+\dot C_{\bf q}(t).
\end{equation}
One gets the normalized analog to Eq.~\eqref{eq:deltachi}, 
\begin{eqnarray}
\Delta\chi_{\bf q}(t)&=&\nonumber\frac{-\dot\gamma}{\langle \varrho^*\varrho\rangle^{(\dot\gamma)}}\int_0^\infty ds\biggl\langle\sigma_{xy}e^{\Omega^{\dagger} s}\nonumber\\&&\sum\limits_{i}(\boldsymbol{\partial}_i+\mathbf{F}_i)\cdot\frac{\partial\varrho^{*}}{\partial\mathbf{r}_i}e^{\Omega^{\dagger} t}\varrho\biggr\rangle.
\end{eqnarray}
Since we are left with an equilibrium average, it is useful to express $\Delta\chi_{\bf q}(t)$ in terms of the transient correlator $\Phi_{\bf q}(t)$ as is done in the following subsection.
\subsection{Zwanzig-Mori Formalism -- FDT Holds at $t=0$}
We use an identity obtained in the Zwanzig-Mori projection operator formalism \cite{Goetze89} (see Eq.~(11) in Ref.~\cite{Fuchs02b}) to find the exact relation 
\begin{eqnarray}
\Delta\chi_{\bf q}(t)&=&\int_0^t dt'\,N_{\bf q}(t-t')\,\Phi_{\mathbf{q}}(t'),\label{eq:gl}\\
N_{\bf q}(t)&=&\frac{-\dot\gamma}{\langle \varrho^*\varrho\rangle^{(\dot\gamma)}}\int_0^\infty ds \biggl\langle\sigma_{xy} e^{\Omega^\dagger s}\nonumber\sum\limits_{i}(\boldsymbol{\partial}_i+\mathbf{F}_i)\cdot\frac{\partial\varrho^{*}}{\partial\mathbf{r}_i}\\&&Qe^{Q\Omega^{\dagger}Qt}Q\,\Omega^{\dagger}{\color{black}\varrho}\biggr\rangle,\label{eq:ZM}
\end{eqnarray}
with $P=\varrho\rangle\langle\varrho^*\varrho \rangle^{-1} \langle\varrho^*$ projecting on a subspace of density fluctuations and $Q=1-P$. $N_{\bf q}(t)$ can also be split into the three contributions according to Eq.~\eqref{eq:dom}, which will be done at the very end only. Eq.~\eqref{eq:gl} could be expected to contain a second term, $\Delta\chi_{\bf q}(t=0)\Phi_{\bf q}(t)$, the static coupling at $t=0$. It vanishes in (\ref{eq:gl}), i.e., $\Delta\chi_{\bf q}(0)=0$; $\Delta\chi_{\bf q}$ is real \cite{Fuchs09}, and second and third term in Eq.~\eqref{eq:dom} cancel at $t=0$. The waiting time derivative vanishes at $t=0$ due to symmetry for arbitrary $f$,
\begin{equation}
\left\langle\sigma_{xy}f^{*}f \right\rangle=0.
\end{equation}
It has thus been shown that the equilibrium FDT is exactly valid at $t=0$.
\subsection{Second Projection Step}\label{sec:CH}
MCT approximations for the function $N_{\bf q}(t)$ directly are not useful because they cannot account for the fact that $N_{\bf q}(t)$ is a fast function. This is achieved with a second projection step following Cichocki and Hess \cite{Cichocki87}, see also Refs.~\cite{Fuchs03,Fuchs09,Krueger09}. The adjoint of the Smoluchowski operator is formally decomposed as
\begin{equation}
\Omega^\dagger=\Omega^i+\Omega^\dagger\varrho\rangle\langle\varrho^*\Omega^\dagger\varrho\rangle^{-1}\langle\varrho^*\Omega^\dagger.\label{eq:CH}
\end{equation}
The function $N_{\bf q}(t)$ is then connected to a new function $n_{\bf q}(t)$ governed by $\Omega^i$.  The following identity can be proven by differentiation,
\begin{eqnarray}
\label{eq:ch}N_{\bf q}(t)/\Gamma_{q}&=& n_{\bf q}(t)-\int_0^t dt'\,N_{\bf q}(t')\,m_{\bf q}(t-t'),\\
n_{\bf q}(t)&=&\frac{-\dot\gamma}{\struc \Gamma_q}\int_0^\infty ds \biggl\langle\sigma_{xy} e^{\Omega^\dagger s}\sum\limits_{i}(\boldsymbol{\partial}_i+\mathbf{F}_i)\nonumber\cdot\\&&\cdot\frac{\partial\varrho^{*}}{\partial\mathbf{r}_i}\,Qe^{Q\Omega^{i}Qt}Q\,\Omega^{\dagger}\varrho\biggr\rangle,\label{eq:mi}\\
m_{\bf q}(t)&=&\frac{1}{\struceq \Gamma_q^2}\left\langle\varrho^*\Omega^\dagger Qe^{Q\Omega^{i}Q t}Q\,\Omega^{\dagger}\varrho\right\rangle.\label{eq:m}
\end{eqnarray}
$\Gamma_q=-\langle \varrho^*\Om\varrho\rangle/\langle \varrho^*\varrho\rangle$ is the initial decay rate. It equals $q^2/S_q$ for the coherent and $q^2$ for the incoherent case. $m_{\bf q}$ is identified as the memory function, which appears in the equation of motion of the transient correlator. It is related to the transient correlator for $q_x=0$ exactly by \cite{Fuchs05}
\begin{equation}
\dot\Phi_{\bf q}(t)+\Gamma_q\left\{ \Phi_{\bf q}(t)+\int_0^t\!\!dt' m_{\bf q}(t-t')\dot\Phi_{\bf q}(t')\right\}=0.\label{eq:mmct}
\end{equation}
We will later be interested in the Laplace transform, $m_{\bf q}(z)=\int_0^\infty dt e^{-zt}m_{\bf q}(t)$, which at $z=0$ is in the limit of slow dynamics given with Eq.~\eqref{eq:mmct} by \cite{Fuchs98}, 
\begin{equation}
m_{\bf q}(z=0)=\Phi_{\bf q}(z=0).\label{eq:mbeiz=0}
\end{equation}
The benefit of the second projection step can now be illuminated by regarding $N_{\bf q}(z)$, which is given via Eq.~(\ref{eq:ch}) by 
\begin{equation}
N_{\bf q}(z)=\frac{n_{\bf q}(z)}{1/\Gamma_{q}+m_{\bf q}(z)}.\label{eq:lap}
\end{equation}
To discuss this, we turn to scaling considerations.
$m_{\bf q}(t)$ decays on  timescale $(\tau^{-1}+|\dot\gamma|)^{-1}$ \cite{Fuchs03}, where $\tau$ is the $\alpha$ relaxation time of the un-sheared system. In the glass, it is formally infinite and we have 
\begin{eqnarray}
\label{eq:mz}
\lim_{\dot\gamma\to0}m_{\bf q}(z=0) =
\left\{\begin{array}{cc}
\mathcal{O}(\tau) & \rm liquid,\\
\mathcal{O}(\frac{1}{|\dot\gamma|}) & \rm glass.
\end{array} \right.
\end{eqnarray}
As will be shown below (Eq.~\eqref{eq:nsch}) by MCT approximations,  $n_{\bf q}(z)$ has the following properties,  
\begin{eqnarray}
\lim_{\dot\gamma\to0}n_{\bf q}(z=0) = 
\left\{\begin{array}{cc}
\mathcal{O}(\dot\gamma^2\tau^2) & \rm liquid,\\ 
\mathcal{O}(1) & \rm glass.
\end{array} \right.
\label{eq:nz}
\end{eqnarray}
The scaling $m_{\bf q}(z=0)\propto|\dot\gamma|^{-1}$ in the glass provides $N_{\bf q}(z=0)$ via Eq.~(\ref{eq:lap}) with an additional power of $\dot\gamma$ compared to $n_{\bf q}(z=0)$,
\begin{eqnarray}
\lim_{\dot\gamma\to0} N_{\bf q}(z=0) =
\left\{\begin{array}{cc}
\mathcal{O}(\dot\gamma^2\tau) & \rm liquid,\\
\mathcal{O}(|\dot\gamma|) & \rm glass.
\end{array} \right.\label{eq:M}
\label{eq:mipowers}
\end{eqnarray}
This is the benefit of the second projection. Note that $N_{\bf q}(z=0)$ corresponds to a rate, which is always small but changes at the glass transition. Because of its smallness, $N_{\bf q}(z=0)$  is difficult to approximate quantitatively.
 
The time integrated violating term $\Delta\chi_{\bf q}$ finally follows at small shear rates with Eq.~\eqref{eq:gl}, 
\begin{eqnarray}
\lim_{\dot\gamma\to0} \Delta\chi_{\bf q}(z=0) =
\left\{\begin{array}{cc}
\mathcal{O}(\dot\gamma^2\tau^2) & \rm liquid,\\
\mathcal{O}(1) & \rm glass.
\end{array} \right.
\label{eq:deltapowers}
\end{eqnarray}
Eq.~\eqref{eq:deltapowers} is in accordance with simulations and with the physically expected property of $\chi_{\bf q}$ to be always finite at $z=0$. The response should not diverge. 

In both the liquid and the glass, the violating term $\Delta\chi_{\bf q}$ is symmetric in $\dot\gamma$, reflecting the fact that fluctuations in $z$- and $y$-direction are independent of the direction of shearing. While $\Delta\chi_{\bf q}$ is analytic in $\dot\gamma$ in the fluid, it is nonanalytic in the glass. 

We can now compare the derived property in Eq.~\eqref{eq:deltapowers} to the approximation in Eq.~\eqref{eq:FDT}, from which we find for glassy states and $\dot\gamma\to0$,
\begin{equation}
\Delta\chi_{\rm f}(z=0)\approx \mathcal{L}\left\{\left.\frac{1}{2}\frac{\partial}{\partial t_w} C_{\rm f}(t,t_w)\right|_{t_w=0}\right\}(z=0)\approx-\frac{1}{2} f_{\rm f},\label{eq:waitz0}
\end{equation}
in accordance with Eq.~\eqref{eq:deltapowers}. $f_{\rm f}$ is the non-ergodicity parameter. In the fluid, Eqs.~\eqref{eq:deltapowers} and \eqref{eq:FDT} also agree which can only be shown numerically, see Fig.~\ref{fig:schem3}. 
\subsection{Markov Approximation -- Long Time FDR}
Using a special projection step, we have shown in the previous subsection that the function $N_{\bf q}(z=0)$ is of order $|\dot\gamma|$ in the glass, i.e., we have reason to assume that $N_{\bf q}(t)$ decays fast in time compared to $\Phi_{\bf q}(t)$ which diverges like $|\dot\gamma|^{-1}$ at $z=0$. With this assumption, Eq.~(\ref{eq:gl}) can be written in Markov approximation using the $\delta$-function, $N_{\bf q}(t)\approx N_{\bf q}(z=0)\,\delta(t)$. For the susceptibility follows
\begin{equation}
\chi_{\bf q}(t)\approx-\frac{\partial}{\partial t} C_{\bf q}(t)+N_{\bf q}(z=0)\Phi_{\bf{q}}(t)\label{eq:markov}.
\end{equation}
We will see in Subsec.~\ref{sec:comparing} that Eq.~\eqref{eq:markov} gives very similar results to Eq.~\eqref{eq:FDT}. According to Eq.~\eqref{eq:markov}, the equilibrium FDT is violated if $N_{\bf q}(z=0)$ is nonzero. For short times, $\dot\gamma t\ll1$, we have $|N_{\bf q}(z=0)|\ll |\dot C_{\bf q}(t)|$ and the equilibrium FDT holds. $N_{\bf q}(z=0)$ is of order $|\dot\gamma|$, see \eqref{eq:M}, $\dot C_{\bf q}(t)$ is of order $\Gamma_q=\mathcal{O}(1)$. For long times, $\dot\gamma t\approx1$, $\dot C_{\bf q}(t)$ is also of order $|\dot\gamma|$, the two terms are comparable in size and the equilibrium FDT is violated. The long time FDR is additionally independent of time, if $C_{\bf q}(t)$ and $\Phi_{\bf q}(t)$ decay exponentially for long times with the same timescale. Approximating the two correlators to be equal and relaxing exponentially in glassy states, 
\begin{equation}
\lim_{t\to\infty} C_{\bf q}(t)\approx \Phi_{\bf q}(t)\approx f_q e^{-a_{\bf q} |\dot\gamma| t},\label{eq:transient}
\end{equation}
we find in the glass,
\begin{equation}
\chi_{\bf q}(t)=\left\{\begin{array}{ll}
-\frac{\partial}{\partial t} C_{\bf q}(t)&\dot\gamma t \ll1,\\
-\left(1+\frac{N_{\bf q}(z=0)}{|\dot\gamma|a_{\bf q}} \right) \frac{\partial}{\partial t} C_{\bf q}(t)& \dot\gamma t=\mathcal{O}(1).
\end{array}
\right.
\end{equation}
This equation is in qualitative agreement with Eq.~\eqref{eq:shandlo}. 
The long time FDR
\begin{equation}
X_{\bf q}(t\to\infty)=1+\frac{N_{\bf q}(z=0)}{|\dot\gamma|a_{\bf q}}\label{eq:FDRmark}
\end{equation}
is time independent and also independent of shear rate for $\dot\gamma\to0$. It is hence non-analytic as pointed out before. As we will see, $N_{\bf q}(z=0)$ is negative in MCT approximations and the FDR is smaller than unity in agreement with Eq.~\eqref{eq:FDT}.
\subsection{FDT Violation Quantitative -- Connection of the two approaches}\label{sec:quant}
We will now perform MCT approximations for the memory function $n_{\bf q}(t)$. It contains two evolution operators, one for the correlation time $t$ and one for the transient time $s$, which entered through the ITT approach. We rewrite Eq.~\eqref{eq:mi} via the identity \eqref{eq:sep} which leads to three contributions for $n_{\bf q}(t)$\begin{subequations}\label{eq:theviol}
\begin{align}
n_{\bf q}(t)=&n_{\bf q}^{(1)}(t)+n_{\bf q}^{(2)}(t)+n_{\bf q}^{(3)}(t)\nonumber,\\
n_{\bf q}^{(1)}(t)=&\frac{-\dot\gamma}{2\struc \Gamma_{q}}\int_0^\infty ds\left\langle\sigma_{xy} e^{\Omega^\dagger s} \Omega^\dagger\varrho^{*}U(t)\Om\varrho\right\rangle,\nonumber\\
=&\frac{\dot\gamma}{2\struc \Gamma_q}\left\langle\sigma_{xy} \varrho^{*}U(t)\Om\varrho\right\rangle,\label{eq:n1}\\
n_{\bf q}^{(2)}(t)=&\frac{\dot\gamma}{2\struc \Gamma_q}\int_0^\infty ds\Bigl\langle \sigma_{xy} e^{\Omega^\dagger s}\nonumber\\& \varrho^{*}\Omega^\dagger U(t)\Om\varrho\Bigr\rangle,\label{eq:n2}\\
n_{\bf q}^{(3)}(t)=&\frac{-\dot\gamma}{2\struc \Gamma_q}\int_0^\infty ds\Bigl\langle \sigma_{xy} e^{\Omega^\dagger s} \nonumber\\&(\Omega^\dagger\varrho^{*}) U(t)\Om\varrho\Bigr\rangle,\label{eq:n3}\\
U(t)=&Qe^{Q\Omega^i Qt}Q.\nonumber
\end{align}
\end{subequations}
The $s$-integration in $n_{\bf q}^{(1)}$ could be done directly as in Eq.~\eqref{eq:chi}. Also $n^{(2)}_{\bf q}(t)$ and $n^{(3)}_{\bf q}(t)$ can now be identified with derivatives with respect to $s$. It is important to note that without identifying these $s$-derivatives, the correct $\dot\gamma$-dependence of $\Delta\chi_{\bf q}(t)$ would not be achieved.  

The detailed MCT approximations for the terms above are shown and evaluated in Appendix~\ref{app:ver}. In the following subsection, we compare the Appendix approach to the main-text approach.
\subsubsection{Connection of the two approaches}\label{sec:comparing}
Let us compare the three terms of $\Delta\chi_{\rm f}$ in Eq.~\eqref{eq:dom} to Eq.~\eqref{eq:theviol}. From the analysis in this Appendix, we have the exact relation for $f=\varrho$, which is more handy in Laplace space,
\begin{equation}
\Delta\chi_{\bf q}(z)=\frac{n_{\bf q}(z)}{1/\Gamma_q+m_{\bf q}(z)}\Phi_{\bf q}(z).
\end{equation}
The waiting time derivative in Eq.~\eqref{eq:dom} is for density fluctuations exactly given by
\begin{equation}
\mathcal{L}\left\{\left.\frac{1}{2}\frac{\partial}{\partial t_w} C_{\bf q}(t,t_w)\right|_{t_w=0} \right\}(z)=\frac{n^{(1)}_{\bf q}(z)}{1/\Gamma_q+m_{\bf q}(z)}\Phi_{\bf q}(z).
\end{equation}
And the sum of the other two terms is exactly given by
\begin{eqnarray}
\mathcal{L}\biggl\{\frac{1}{2}\left[\dot C_{\bf q}(t)-\dot \Phi_{\bf q}(t)\right]+\Delta\chi_{\bf q}^{(3)}(t)\biggr\}(z)\nonumber=\\ \frac{n^{(2)}_{\bf q}(z)+n^{(3)}_{\bf q}(z)}{1/\Gamma_q+m_{\bf q}(z)}\Phi_{\bf q}(z).\label{eq:summe}
\end{eqnarray}
With the Markov approximation in Eq.~\eqref{eq:markov}, these are simplified to (note that $m_{\bf q}(z=0)\propto|\dot\gamma|^{-1}\gg 1/\Gamma_q$)
\begin{subequations}\label{eq:nu}
\begin{align}
\frac{1}{2}\left.\frac{\partial}{\partial t_w} C_{\bf q}(t,t_w)\right|_{t_w=0}\approx\frac{n^{(1)}_{\bf q}(z=0)}{m_{\bf q}(z=0)}\Phi_{\bf q}(t),\label{eq:vergl}\\
\mathcal{L}^{-1}\left\{\eqref{eq:summe}\right\}(t)\approx\frac{n^{(2)}_{\bf q}(z=0)+n^{(3)}_{\bf q}(z=0)}{m_{\bf q}(z=0)}\Phi_{\bf q}(t).\label{eq:vergl2}
\end{align}
\end{subequations}
with $n^{(i)}_{\bf q}(z=0)/m_{\bf q}(z=0)\propto |\dot\gamma|$ as $\dot\gamma\to0$, see Eq.~\eqref{eq:nz}. From Eq.~\eqref{eq:lti}, which we consider very accurate, the waiting time derivative in glassy states is for long times equal to $\dot \Phi_{\bf q}(t)$.
We conclude that Eqs.~\eqref{eq:lti} and \eqref{eq:vergl} are in qualitative agreement, if the transient correlator decays exponentially for long times with timescale $\propto |\dot\gamma|^{-1}$. This holds well \cite{Fuchs03} and also has been used in the main text, see Eq.~\eqref{eq:corr1}. 

\begin{figure}
\begin{center}\includegraphics[width=1\linewidth]{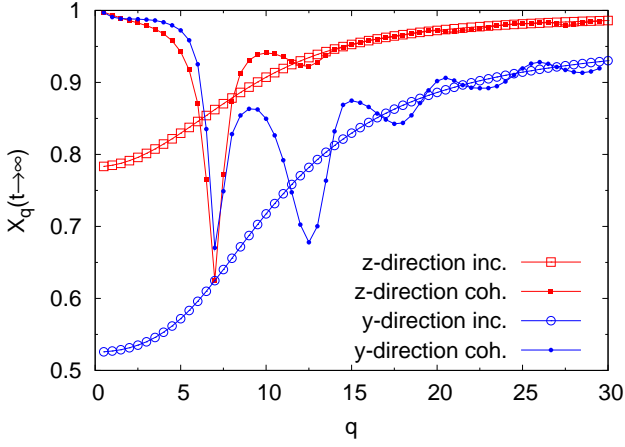}\end{center}
\caption{Long time FDR as function of wavevector $q$ for density fluctuations in $z$- and $y$-direction in the limit of small shear rates. The packing fraction is the critical one ($\varepsilon=0$). \label{fig:X}}
\end{figure}

In Fig.~\ref{fig:memories}, we show the quantitative comparison of the functions at $z=0$, see Appendix \ref{app:ver} for details on the MCT approximations for the terms \eqref{eq:theviol} and their numerical evaluation. We see, that the MCT-estimates compare quite well to the prediction of Eq.~\eqref{eq:lti}. Fig.~\ref{fig:X} shows the long time FDR as function of $\bf q$ as calculated by the MCT approximations for $n_{\bf q}$ and Eq.~\eqref{eq:FDRmark}. We show only the contribution of the first term, i.e., $n_{\bf q}(z=0)=n^{(1)}_{\bf q}(z=0)$. Also, in Eq.~\eqref{eq:FDRmark}, the difference between stationary and transient correlators is neglected since it follows with Eq.~\eqref{eq:transient}. According to our analysis in the main text, Eq.~\eqref{eq:lti}, these two simplifications would yield $X=\frac{1}{2}$ for all $q$. We see that the FDR evaluated from \eqref{eq:FDRmark} depends rather strongly on wavevector, but given the complexity of the involved functions, the result is still satisfying.

\section{MCT-Approximations for $n_{\bf q}(t)$}\label{app:ver}
Here we show the detailed approximations for the formally exact expressions for $n_{\bf q}$ in \eqref{eq:theviol} by projection onto densities. This physical approximation amounts to assuming that these are the only slow variables, sufficient to describe the relaxation of the local structure in the glassy regime. 
\subsection{Coherent Case}
For coherent density fluctuations, the time dependent pair density projector is given by \cite{Fuchs09}
\begin{equation}
P_2(t)=\sum_{\bf k>p}\frac{\varrho_{\vct{k}(t)}\varrho_{\vct{p}(t)}\rangle\langle \varrho^*_{\vct{k}(t)}\varrho^*_{\vct{p}(t)}}{N^2 S_{\vct{k}(t)} S_{\vct{p}(t)}}.
\end{equation}
The memory function $n_{\bf q}^{(1)}$ has only one time evolution operator and can hence be approximated via `standard' routes with the projector $P_2(t)$ \cite{Fuchs09},
\begin{equation}
n_{\bf q}^{(1)}(t)\approx\frac{\dot\gamma}{2NS^{(\dot\gamma)}_{\bf q} \Gamma_{q}}\left\langle\sigma_{xy} \varrho_{\bf{q}}^{*}QP_2(-t)U(t)P_2\Omega^{\dagger}\varrho_{\bf q}\right\rangle.\label{eq:mi2}\\
\end{equation}
For the appearing time dependent four point correlation function, the factorization approximation is used
\begin{eqnarray}
\langle\varrho^*_{\vct{k'}(-t)}\varrho^*_{\vct{k'}(-t)-\vct{q}} U(t)\varrho_{\vct{k}}\varrho_{\vct{k-q}}\rangle\nonumber\approx\\ N^2 S_{k(-t)} S_{k(-t)-q} \Phi_{\vct{k}(-t)}(t)\Phi_{\vct{k}(-t)-\vct{q}}(t)\delta_{\vct{k},\vct{k}'}.
\end{eqnarray}
At the left hand side appears the expression
\begin{eqnarray}
V^{(1)}_{\bf qk}&=&\frac{\left\langle\sigma_{xy}\varrho_{\bf q}^* Q\varrho_{\bf k}\varrho_{\bf q-k} \right\rangle}{N S_q}\nonumber\\&\approx& k_x(k_y-q_y)\frac{S'_{q-k}}{q-k}S_{k}+k_xk_y\frac{S'_{k}}{k}S_{q-k}.\label{eq:vertex1}
\end{eqnarray}
On the right hand side, we have the standard vertex \cite{Fuchs09}
\begin{equation}
V^{(2)}_{\bf qk}=\frac{\left\langle \varrho_{\bf k} \varrho_{\bf q-k}Q\Omega^\dagger  \varrho_{\bf q}\right\rangle}{N S_{k-q}S_k}\approx{\bf q}\cdot (({\bf k}-{\bf q})nc_{k-q}+{\bf k}nc_{k}).\label{eq:vertex2}
\end{equation}
In the derivation of Eqs. \eqref{eq:vertex1} and  \eqref{eq:vertex2}, the convolution approximation for the static three point correlation function was used \cite{Goetze91}, 
\begin{equation}
\left\langle \varrho_{\bf q}\varrho_{\bf k-q} \varrho_{\bf k}\right\rangle\approx N S_q\,S_{k-q}\,S_{k}.
\end{equation}
The treatment of $n_{\bf q}^{(2)}$ and $n_{\bf q}^{(3)}$ is more involved. We first separate the time evolution operator in $t$ with pair projectors to get, 
\begin{eqnarray}
n_{\bf q}^{(2)}(t)&\approx&\frac{\dot\gamma}{2NS^{(\dot\gamma)}_{\bf q} \Gamma_q}\int_0^\infty ds\Bigl\langle\sigma_{xy}e^{\Omega^\dagger s}\nonumber\\&&\varrho_{\bf{q}}^{*}\Omega^\dagger \,Q\,P_2 U(t) P_2(t) Q\Omega^{\dagger}{\color{black}\varrho_{\vct{q}}}\Bigr\rangle,\label{eq:erste}\\
n_{\bf q}^{(3)}(t)&\approx&\frac{-\dot\gamma}{2NS^{(\dot\gamma)}_{\bf q} \Gamma_q}\int_0^\infty ds\Bigl\langle\sigma_{xy} e^{\Omega^\dagger s}\nonumber\\&&(\Omega^\dagger\varrho_{\bf{q}}^{*}) \,Q\,P_2 U(t) P_2(t) Q\Omega^{\dagger}{\color{black}\varrho_{\vct{q}}}\Bigr\rangle.\label{eq:zweite}
\end{eqnarray}
After doing this, we are on the left hand side of the projectors left with the two respective expressions,
\begin{equation}
\left\langle\sigma_{xy} e^{\Omega^\dagger s}\varrho_{\bf{q}}^{*}\Omega^\dagger \,Q\,\varrho_{\bf q-k}\varrho_{\bf k}\right\rangle \label{eq:erste1},
\end{equation}
and
\begin{equation}
\left\langle\sigma_{xy} e^{\Omega^\dagger s}(\Omega^\dagger\varrho_{\bf{q}}^{*}) \,Q\,\varrho_{\bf q-k}\varrho_{\bf k}\right\rangle.\label{eq:zweite2}
\end{equation}
Writing $Q$ as $1-P$, we realize that the term containing $P$ is identical for both terms (they are real),
\begin{equation}
\left\langle\sigma_{xy} e^{\Omega^\dagger s}(\Omega^\dagger\varrho_{\bf{q}}^{*}) \,\varrho_{\bf q}\right\rangle\frac{1}{NS_q}\left\langle\varrho_{\bf q}^*\varrho_{\bf q-k}\varrho_{\bf k}\right\rangle,                            \label{eq:cancel}
\end{equation}
with opposite sign. These terms cancel each other. We are left with the two expressions,
\begin{equation}
\left\langle\sigma_{xy} e^{\Omega^\dagger s}\varrho_{\bf{q}}^{*}\Omega^\dagger \varrho_{\bf q-k}\varrho_{\bf k}\right\rangle\label{eq:in2},
\end{equation}
and
\begin{equation}
\left\langle\sigma_{xy} e^{\Omega^\dagger s}(\Omega^\dagger\varrho_{\bf{q}}^{*}) \varrho_{\bf q-k}\varrho_{\bf k}\right\rangle.           \label{eq:in3}                 \end{equation}
There is in principal more than one option to treat these terms, but we will argue that only one option is applicable. The standard way, i.e., the  usage of $P_2$ right and left of the time evolution operator is not preferable since it would not preserve the derivative with respect to $s$. As already mentioned, this derivative is necessary for the correct $\dot\gamma$-dependence. That is why we chose to use the triple densities projector $P_3$,
\begin{equation}
P_3(t)=\sum_{\bf k>p>n}\frac{\varrho_{\vct{k}(t)}\varrho_{\vct{p}(t)}\varrho_{\vct{n}(t)}\rangle\langle \varrho^*_{\vct{k}(t)}\varrho^*_{\vct{p}(t)}\varrho^*_{\vct{n}(t)}}{N^3 S_{\vct{k}(t)} S_{\vct{p}(t)}S_{\vct{n}(t)}}.\label{eq:triple}
\end{equation}
Eq. \eqref{eq:in2} is written as
\begin{equation} \left\langle\sigma_{xy} e^{\Omega^\dagger s}\varrho_{\bf{q}}^{*}\Omega^\dagger \varrho_{\bf q-k}\varrho_{\bf k}\right\rangle\approx \left\langle\sigma_{xy} P_3(-s) e^{\Omega^\dagger s}\varrho_{\bf{q}}^{*}\Omega^\dagger \varrho_{\bf q-k}\varrho_{\bf k}\right\rangle.\label{eq:A14}
\end{equation}
We have to demand that the wave-vectors in the triple projector take the values of the wave-vectors on the right hand side. Due to this constraint, the summation in Eq.~\eqref{eq:triple} contains only one term and no counting factor appears. The left hand side is the Vertex $V^{(1)}_{\bf qk}$ in Eq. \eqref{eq:vertex1} for $s$--dependent wave-vectors (the projector $Q$ does not make any difference in \eqref{eq:vertex1}).
The appearing six point $s$-dependent correlation function is approximated as,
\begin{eqnarray}
\frac{\langle\varrho_{\vct{q}}^*\varrho^*_{\vct{k}(-s)}\varrho^*_{\vct{k}(-s)-\vct{q}} e^{\Omega^\dagger s}\varrho_{\vct{q}} \Omega^\dagger\varrho_{\vct{k}}\varrho_{\vct{k-q}}\rangle}{N^3 S_q S_{k(-s)-q}S_{k(-s)}}\nonumber\approx\\\Phi_{\vct{q}}(s)\left(\left[\frac{\partial}{\partial s}-{\bf k}(-s)\cdot\frac{\partial}{\partial {\bf k}}\right]\Phi_{{\bf k}(-s)}(s)\Phi_{{\bf k}(-s)-{\bf q}}(s)\right)\label{eq:sderivative}.\nonumber
\end{eqnarray}
This approximation rests on the observation that the operator $\Omega^\dagger$ acts as an $s$ derivative. $\Phi_{{\bf k}(-s)}(s)$ and $\Phi_{{\bf k}(-s)-{\bf q}}(s)$ depend on $s$ via the decay of the correlator and via the $s$--dependent  wave-vectors. Since $\Omega^\dagger$ in (\ref{eq:sderivative}) represents the $s$--derivative with respect to correlator dynamics, we have to subtract the change in $s$ due to the change of the wave-vectors. 
The term in Eq. \eqref{eq:in3} is treated analogously with the triple density projector. Here, the approximation for the appearing six point correlation function is more straight forward, since $\Phi_{\bf q}(s)$ has no wavevector advection,
\begin{eqnarray}
\frac{\langle\varrho_{\vct{q}}^*\varrho^*_{\vct{k}(-s)}\varrho^*_{\vct{k}(-s)-\vct{q}} e^{\Omega^\dagger s}(\Omega^\dagger\varrho_{\vct{q}}) \varrho_{\vct{k}}\varrho_{\vct{k-q}}\rangle}{N^3\,S_{q}\,S_{{k}(-s)-{q}}\,S_{k(-s)}}\nonumber\approx\\ \left(\frac{\partial}{\partial s}\Phi_{\vct{q}}(s)\right)\Phi_{{\bf k}(-s)}(s)\Phi_{{\bf k}(-s)-{\bf q}}(s)\label{eq:sderivative2}.\nonumber
\end{eqnarray}
Collecting the terms, we finally find the following expressions, where $\bar n_{\bf q}^{(2)}$ and $\bar n_{\bf q}^{(3)}$ denote the functions without the terms in Eq.~\eqref{eq:cancel},    
\begin{widetext}
\begin{subequations}\label{eq:violap}
\begin{eqnarray}
n^{(1)}_{\bf q}(t)&=&\frac{\dot\gamma S_q^2}{4n q^2S_{\bf q}^{(\dot\gamma)}}\int\frac{d^3k}{(2\pi)^3}V^{(1)}_{{\bf qk}(-t)} V^{(2)}_{\bf qk}\Phi_{{\bf k}(-t)}(t)\Phi_{{\bf k}(-t)-{\bf q}}(t),\label{eq:n1ap}\\
\bar n^{(2)}_{\bf q}(t)&=&\frac{\dot\gamma S_q^2}{4n q^2S_{\bf q}^{(\dot\gamma)}}\int_0^\infty ds\int\frac{d^3k}{(2\pi)^3}V^{(1)}_{{\bf qk}(-s)} V^{(2)}_{{\bf qk}(t)}\Phi_{\bf q}(s)\left(\left[\frac{\partial}{\partial s}-{\bf k}'(-s)\cdot\frac{\partial}{\partial {\bf k}}\right]\Phi_{{\bf k}(-s)}(s)\Phi_{{\bf k}(-s)-{\bf q}}(s)\right)\Phi_{{\bf k}}(t)\Phi_{{\bf k}-{\bf q}}(t),\nonumber\\\label{eq:n2ap}\\
\bar n^{(3)}_{\bf q}(t)&=&\frac{-\dot\gamma S_q^2}{4n q^2S_{\bf q}^{(\dot\gamma)}}\int_0^\infty ds\int\frac{d^3k}{(2\pi)^3}V^{(1)}_{{\bf qk}(-s)} V^{(2)}_{{\bf qk}(t)}\left(\frac{\partial}{\partial s}\Phi_{\bf q}(s)\right)\Phi_{{\bf k}(-s)}(s)\Phi_{{\bf k}(-s)-{\bf q}}(s)\Phi_{{\bf k}}(t)\Phi_{{\bf k}-{\bf q}}(t),\label{eq:n3ap}\\
V^{(1)}_{\bf qk}&=&k_x(k_y-q_y)\frac{S'_{q-k}}{q-k}S_{k}+k_xk_y\frac{S'_{k}}{k}S_{q-k},\hspace{1cm}
V^{(2)}_{\bf qk}={\bf q}\cdot (({\bf k}-{\bf q})nc_{k-q}+{\bf k}nc_{k}).\nonumber
\end{eqnarray}
\end{subequations}
\end{widetext}
With $q-k=|{\bf q-k}|$ and ${\bf k}(t)={\bf k}-{\bf k}\cdot\kap \,t$ as before.
$c_k$ is the equilibrium direct correlation function connected to the structure factor via the Ornstein-Zernicke equation
$S_k=1/(1-nc_k)$ \cite{Hansen}. 
From the expressions in \eqref{eq:violap} one can now see the earlier proposed properties, Eq.~(\ref{eq:nz}). The function $n_{\bf q}$ can schematically be written
\begin{equation}
n_{\bf q}(t)=a\,\dot\gamma^2 \,t f(t)+b\, \int_0^\infty ds\, \dot\gamma^2 \,s\, \,  \frac{\partial}{\partial s}g(s)\,h(t).\label{eq:nsch}
\end{equation}
The first term in \eqref{eq:nsch} corresponds to $n_{\bf q}^{(1)}$, the second term to $n_{\bf q}^{(2)}$ and $n_{\bf q}^{(3)}$. $f(t), g(t), h(t)$ are functions of $t/\tau$ in the liquid and $|\dot\gamma| t$ in the glass. Eq.~(\ref{eq:nz}) follows.  The fact that the terms in \eqref{eq:nsch}  start linearly with $t$ and $s$ respectively comes because $V^{(2)}$ in Eq.~\eqref{eq:violap} is symmetric in $k_x$,
$V^{(2)}=V^{(2)}(k_x^2)$, and because $V^{(1)}$ at time $t=0$ (or $s=0$) is
anti-symmetric in $k_x$, $V^{(1)}(-k_x)=-V^{(1)}(k_x)$, and the
property $n_{\bf q}^{(1)}(t\to 0)\sim \dot\gamma^2 t$ follows after integration over $d^3k$. The linear increase with time follows for example from
$k_y(t)=k_y-\dot\gamma t k_x$. 

For the numerical evaluation of Eq.~(\ref{eq:violap}), the transient correlator $\Phi_{\bf q}(t)$ and the static structure factors  $S_q$ and $S_{\bf q}^{(\dot\gamma)}$ are needed. As a purely technical simplification, we use the isotropic approximation \cite{Fuchs03}, which reads for long times in glassy states
\begin{equation}
\Phi_{\bf q}(t)\approx\Phi_q(t)=f_q\,e^{-c\frac{h_q}{f_q} |\dot\gamma| t},\label{eq:corr}
\end{equation}
with the non ergodicity parameter $f_q$ and the amplitude $h_q$. The parameter $c$ can be derived from a microscopic analysis, we use $c=3$ \cite{Fuchs03}.
For the static equilibrium structure factor, we use the Percus-Yevick closure \cite{Hansen}, and approximate $S_q=S_{\bf q}^{(\dot\gamma)}$, which holds well at small shear rates \cite{Berthier02}, although the structure is nonanalytic \cite{Henrich07}.  In the limit of small shear rates, the contribution of the short time decay of the correlators to the above expressions vanishes. The above expressions are evaluated using spherical coordinates with grid $k_{max}=50$, $\Delta k=0.05$ and $\Delta \theta=\Delta\phi=\pi/40$ or smaller. The time grid in both $t$ and $s$ was $\dot\gamma t=2^{i/4}/10^{10}$, starting from $i=90$ corresponding to $\dot\gamma t\approx 6\,\,10^{-4}$. The results are included in Figs.~\ref{fig:memories} and \ref{fig:X}.
\subsection{Incoherent Case}
From now on we denote incoherent functions with superscript $s$. The terms in Eq.~\eqref{eq:theviol} for the incoherent case, $f=\varrho_{\bf q}^s$ are approximated similarly to the coherent analogs, using the pair density projector \cite{incoherent}
\begin{equation}
P^s_2(t)=\sum_{\vct{p},\vct{k}}\frac{\varrho_{{\bf p}(t)}^s\varrho_{{\bf k}(t)}\rangle\langle\varrho^{s*}_{\vct{p}(t)}\varrho^*_{\vct{k}(t)}}{N S_{k(t)}}. \label{eq:pairden}
\end{equation}
The approximation for $n_{\bf q}^{(s,1)}(t)$ is then straight forward following Eqs. (\ref{eq:mi2}-\ref{eq:vertex1}). Regarding the vertex, there occur simplifications, 
\begin{equation}                                                                                                                                            V^{(s,1)}_{\bf qk}=\left\langle\sigma_{xy}\varrho_{\bf q}^{s*} Q^s\varrho_{\bf k}\varrho^s_{\bf q-k} \right\rangle= k_xk_y\frac{S'_{k}}{k}.\label{eq:vertex1
s}                                                                                                                                                          \end{equation}
The right hand side of the vertex reads
\begin{equation}
V_{\vct{q}\vct{k}}^{(s,2)}=\frac{\left\langle\varrho^{s}_{\bf k-q}\varrho_{\bf -k}\,Q^s\,\Omega^\dagger_e\varrho_{\bf q}^s\right\rangle}{S_k}=\vct{k}\cdot\vct{q}\,n\,c_k^{s}.
\end{equation}
For the memory functions $n_{\bf q}^{(s,2)}(t)$ and $n_{\bf q}^{(s,3)}(t)$ we first use the projector $P^s_2$ according to Eqs. \eqref{eq:erste} and \eqref{eq:zweite}. We note that the two terms corresponding to Eq.~\eqref{eq:cancel} vanish in this case independently. We arrive at the expressions equivalent to Eqs. \eqref{eq:in2} and \eqref{eq:in3}, reading $\langle\sigma_{xy} \exp[{\Omega^\dagger s}]\varrho_{\bf{q}}^{s*}\Omega^\dagger \varrho^s_{\bf q-k}\varrho_{\bf k}\rangle$ and $\langle\sigma_{xy} \exp[{\Omega^\dagger s}](\Omega^\dagger\varrho_{\bf{q}}^{s*}) \varrho^s_{\bf q-k}\varrho_{\bf k}\rangle$.
We use the triple density projector as before,
\begin{equation}
P^s_3(t)\approx\sum_{\bf k>p>n}\frac{\varrho^s_{\vct{k}(t)}\varrho^s_{\vct{p}(t)}\varrho_{ \vct{n}(t)}\rangle\langle \varrho^s_{\vct{k}(t)}\varrho^s_{\vct{p
}(t)}\varrho_{\vct{n}(t)}}{N S_{{n}(t)}},\label{eq:tripleinc}                                                                                               \end{equation}
according to Eq.~\eqref{eq:A14}. The discussions around Eqs.~\eqref{eq:sderivative} and \eqref{eq:sderivative2} and the approximations for the six-point functions hold similarly. We arrive at
\begin{widetext}
\begin{subequations}\label{eq:ninco}
\begin{eqnarray}
n^{(s,1)}_{\bf q}(t)&=&\frac{\dot\gamma}{2n q^2}\int\frac{d^3k}{(2\pi)^3}V^{(s,1)}_{{\bf qk}(-t)} V^{(s,2)}_{\bf qk}\Phi_{{\bf k}(-t)}(t)\Phi^{s}_{{\bf k}(-t)-{\bf q}}(t),\label{eq:n1apinc}\\
n^{(s,2)}_{\bf q}(t)&=&\frac{\dot\gamma }{2n q^2}\int_0^\infty ds\int\frac{d^3k}{(2\pi)^3}V^{(s,1)}_{{\bf qk}(-s)} V^{(s,2)}_{{\bf qk}(t)}\Phi^{s}_{\bf q}(s)\left(\left[\frac{\partial}{\partial s}-{\bf k}'(-s)\cdot\frac{\partial}{\partial {\bf k}}\right]\Phi_{{\bf k}(-s)}(s)\Phi^s_{{\bf k}(-s)-{\bf q}}(s)\right)\Phi_{{\bf k}}(t)\Phi^s_{{\bf k}-{\bf q}}(t),\nonumber\\\label{eq:n2apinc}\\
n^{(s,3)}_{\bf q}(t)&=&\frac{-\dot\gamma}{2n q^2}\int_0^\infty ds\int\frac{d^3k}{(2\pi)^3}V^{(s,1)}_{{\bf qk}(-s)} V^{(s,2)}_{{\bf qk}(t)}\left(\frac{\partial}{\partial s}\Phi^{s}_{\bf q}(s)\right)\Phi_{{\bf k}(-s)}(s)\Phi^s_{{\bf k}(-s)-{\bf q}}(s)\Phi_{{\bf k}}(t)\Phi^{s}_{{\bf k}-{\bf q}}(t),\label{eq:n3apinc}\\
V^{(s,1)}_{\bf qk}&=&\left\langle\sigma_{xy}\varrho_{\bf q}^{s*} Q^s\varrho_{\bf k}\varrho^s_{\bf q-k} \right\rangle= k_xk_y\frac{S'_{k}}{k},\hspace{1cm} V_{\vct{q}\vct{k}}^{(s,2)}=\frac{\left\langle\varrho^{s}_{\bf k-q}\varrho_{\bf -k}\,Q^s\,\Omega^\dagger_e\varrho_{\bf q}^s\right\rangle}{S_k}=\vct{k}\cdot\vct{q}\,n\,c_k^{s}.\nonumber
\end{eqnarray}
\end{subequations}
\end{widetext}
With $c_q^{s}=\langle \varrho^{s*}_{q}\varrho_{q}\rangle/(n S_q)$
 \cite{Hansen}. We evaluate these expressions numerically, using the approximation \eqref{eq:corr} for the coherent correlator and a similar approximation for the incoherent one, motivated by the solution for the correlator near the critical plateau \cite{incoherent,Krueger09b}. We write for long times in glassy states
\begin{equation}
\Phi^{s}_{\bf q}(t)\approx\Phi^{s}_{ q}(t)=f_q^{s}\exp\left( -\frac{h_q^{s}}{f_q^{s}} c\,|\dot\gamma| t\right),\label{eq:corrinc}
\end{equation}
and again $c=3$. We consider the case where the tagged particle has the same size as the bath particles, for which $c_q^s=(S_q-1)/(nS_q)$ holds.
We use the grid $k_{max}=25$, $\Delta k=0.2$, $\Delta \theta=\pi/40$ and $\Delta \phi=\pi/20$ ($z$-direction) and $\Delta \theta=\pi/20$ and $\Delta \phi=\pi/64$ ($y$-direction) or smaller, the time is discretize as in the coherent case.

The results are included in Figs.~\ref{fig:memories} and \ref{fig:X} and again allow to conclude qualitative agreement with the approach of the main text, with quantitative differences arising from the involved MCT approximations.
 
\section{Simulation details}\label{sec:detsimu}
The shear step model was simulated as follows: The one dimensional random walk within the well was discretized. At each time step (with $\Delta t$ the length of the time step), the particle position propagates by the step length  $s$, $x(t+\Delta t)=x(t)+s$. $s$ was determined by 
\begin{equation}
s=\sum_{i=1}^{12} r_i-6,\label{eq:distr}
\end{equation}
with $r_i$ random numbers (0,1). This gives a Gaussian distribution of width $\sigma=1$. We used $a=4$ for the width of the well. If $x(t+\Delta t)$ lies outside the well, the step is rejected, i.e., the particle stays at its position. For the case with external force $F$, the distribution in Eq.~\eqref{eq:distr} was shifted by $5/100$, corresponding to $F=\frac{k_BT}{10 \sigma}$ with $\sigma^2=2 D_0 \Delta t$. With $\sigma=1$ and $a=4$, the deviation of the linear response result is of the order of 1 percent.
\end{appendix}


\begin{thebibliography}{81}
\expandafter\ifx\csname natexlab\endcsname\relax\def\natexlab#1{#1}\fi
\expandafter\ifx\csname bibnamefont\endcsname\relax
  \def\bibnamefont#1{#1}\fi
\expandafter\ifx\csname bibfnamefont\endcsname\relax
  \def\bibfnamefont#1{#1}\fi
\expandafter\ifx\csname citenamefont\endcsname\relax
  \def\citenamefont#1{#1}\fi
\expandafter\ifx\csname url\endcsname\relax
  \def\url#1{\texttt{#1}}\fi
\expandafter\ifx\csname urlprefix\endcsname\relax\def\urlprefix{URL }\fi
\providecommand{\bibinfo}[2]{#2}
\providecommand{\eprint}[2][]{\url{#2}}

\bibitem[{\citenamefont{Einstein}(1905)}]{Einstein05}
\bibinfo{author}{\bibfnamefont{A.}~\bibnamefont{Einstein}},
  \bibinfo{journal}{Annalen der Physik} \textbf{\bibinfo{volume}{17}},
  \bibinfo{pages}{459} (\bibinfo{year}{1905}).

\bibitem[{\citenamefont{Kubo et~al.}(1985)\citenamefont{Kubo, Toda, and
  Hashitsume}}]{Kubo}
\bibinfo{author}{\bibfnamefont{R.}~\bibnamefont{Kubo}},
  \bibinfo{author}{\bibfnamefont{M.}~\bibnamefont{Toda}}, \bibnamefont{and}
  \bibinfo{author}{\bibfnamefont{N.}~\bibnamefont{Hashitsume}},
  \emph{\bibinfo{title}{Statistitical Physics 2}}
  (\bibinfo{publisher}{Springer}, \bibinfo{address}{Berlin},
  \bibinfo{year}{1985}).

\bibitem[{\citenamefont{Nyquist}(1928)}]{Nyquist28}
\bibinfo{author}{\bibfnamefont{H.}~\bibnamefont{Nyquist}},
  \bibinfo{journal}{Phys. Rev.} \textbf{\bibinfo{volume}{32}},
  \bibinfo{pages}{110} (\bibinfo{year}{1928}).

\bibitem[{\citenamefont{Crisanti and Ritort}(2003)}]{Crisanti03}
\bibinfo{author}{\bibfnamefont{A.}~\bibnamefont{Crisanti}} \bibnamefont{and}
  \bibinfo{author}{\bibfnamefont{F.}~\bibnamefont{Ritort}},
  \bibinfo{journal}{J. Phys. A} \textbf{\bibinfo{volume}{36}},
  \bibinfo{pages}{R181} (\bibinfo{year}{2003}).

\bibitem[{\citenamefont{Risken}(1984)}]{Risken}
\bibinfo{author}{\bibfnamefont{H.}~\bibnamefont{Risken}},
  \emph{\bibinfo{title}{The Fokker-Planck Equation}}
  (\bibinfo{publisher}{Springer}, \bibinfo{address}{Berlin},
  \bibinfo{year}{1984}).

\bibitem[{\citenamefont{Agarwal}(1972)}]{agarwal72}
\bibinfo{author}{\bibfnamefont{G.~S.} \bibnamefont{Agarwal}},
  \bibinfo{journal}{Z. Physik} \textbf{\bibinfo{volume}{252}},
  \bibinfo{pages}{25} (\bibinfo{year}{1972}).

\bibitem[{\citenamefont{Speck and Seifert}(2006)}]{Speck06}
\bibinfo{author}{\bibfnamefont{T.}~\bibnamefont{Speck}} \bibnamefont{and}
  \bibinfo{author}{\bibfnamefont{U.}~\bibnamefont{Seifert}},
  \bibinfo{journal}{Europhys. Lett.} \textbf{\bibinfo{volume}{74}},
  \bibinfo{pages}{391} (\bibinfo{year}{2006}).

\bibitem[{\citenamefont{Blickle et~al.}(2007)\citenamefont{Blickle, Speck,
  Lutz, Seifert, and Bechinger}}]{Blickle07}
\bibinfo{author}{\bibfnamefont{V.}~\bibnamefont{Blickle}},
  \bibinfo{author}{\bibfnamefont{T.}~\bibnamefont{Speck}},
  \bibinfo{author}{\bibfnamefont{C.}~\bibnamefont{Lutz}},
  \bibinfo{author}{\bibfnamefont{U.}~\bibnamefont{Seifert}}, \bibnamefont{and}
  \bibinfo{author}{\bibfnamefont{C.}~\bibnamefont{Bechinger}},
  \bibinfo{journal}{Phys. Rev. Lett.} \textbf{\bibinfo{volume}{98}},
  \bibinfo{pages}{210601} (\bibinfo{year}{2007}).

\bibitem[{\citenamefont{Berthier et~al.}(2000)\citenamefont{Berthier, Barrat,
  and Kurchan}}]{Berthier99}
\bibinfo{author}{\bibfnamefont{L.}~\bibnamefont{Berthier}},
  \bibinfo{author}{\bibfnamefont{J.-L.} \bibnamefont{Barrat}},
  \bibnamefont{and} \bibinfo{author}{\bibfnamefont{J.}~\bibnamefont{Kurchan}},
  \bibinfo{journal}{Phys. Rev. E} \textbf{\bibinfo{volume}{61}},
  \bibinfo{pages}{5464} (\bibinfo{year}{2000}).

\bibitem[{\citenamefont{Berthier and Barrat}(2002{\natexlab{a}})}]{Berthier02}
\bibinfo{author}{\bibfnamefont{L.}~\bibnamefont{Berthier}} \bibnamefont{and}
  \bibinfo{author}{\bibfnamefont{J.-L.} \bibnamefont{Barrat}},
  \bibinfo{journal}{J. Chem. Phys.} \textbf{\bibinfo{volume}{116}},
  \bibinfo{pages}{6228} (\bibinfo{year}{2002}{\natexlab{a}}).

\bibitem[{\citenamefont{Berthier and
  Barrat}(2002{\natexlab{b}})}]{Berthier02prl}
\bibinfo{author}{\bibfnamefont{L.}~\bibnamefont{Berthier}} \bibnamefont{and}
  \bibinfo{author}{\bibfnamefont{J.-L.} \bibnamefont{Barrat}},
  \bibinfo{journal}{Phys. Rev. Lett.} \textbf{\bibinfo{volume}{89}},
  \bibinfo{pages}{095702} (\bibinfo{year}{2002}{\natexlab{b}}).

\bibitem[{\citenamefont{Barrat and Berthier}(2000)}]{Barrat00}
\bibinfo{author}{\bibfnamefont{J.-L.} \bibnamefont{Barrat}} \bibnamefont{and}
  \bibinfo{author}{\bibfnamefont{L.}~\bibnamefont{Berthier}},
  \bibinfo{journal}{Phys. Rev. E} \textbf{\bibinfo{volume}{63}},
  \bibinfo{pages}{012503} (\bibinfo{year}{2000}).

\bibitem[{\citenamefont{O\char39{}Hern
  et~al.}(2004)\citenamefont{O\char39{}Hern, Liu, and Nagel}}]{Ohern04}
\bibinfo{author}{\bibfnamefont{C.~S.} \bibnamefont{O\char39{}Hern}},
  \bibinfo{author}{\bibfnamefont{A.~J.} \bibnamefont{Liu}}, \bibnamefont{and}
  \bibinfo{author}{\bibfnamefont{S.~R.} \bibnamefont{Nagel}},
  \bibinfo{journal}{Phys. Rev. Lett.} \textbf{\bibinfo{volume}{93}},
  \bibinfo{pages}{165702} (\bibinfo{year}{2004}).

\bibitem[{\citenamefont{Haxton and Liu}(2007)}]{Haxton07}
\bibinfo{author}{\bibfnamefont{T.~K.} \bibnamefont{Haxton}} \bibnamefont{and}
  \bibinfo{author}{\bibfnamefont{A.~J.} \bibnamefont{Liu}},
  \bibinfo{journal}{Phys. Rev. Lett.} \textbf{\bibinfo{volume}{99}},
  \bibinfo{pages}{195701} (\bibinfo{year}{2007}).

\bibitem[{\citenamefont{Zamponi et~al.}(2005)\citenamefont{Zamponi, Ruocco, and
  Angelani}}]{Zamponi05}
\bibinfo{author}{\bibfnamefont{F.}~\bibnamefont{Zamponi}},
  \bibinfo{author}{\bibfnamefont{G.}~\bibnamefont{Ruocco}}, \bibnamefont{and}
  \bibinfo{author}{\bibfnamefont{L.}~\bibnamefont{Angelani}},
  \bibinfo{journal}{Phys. Rev. E} \textbf{\bibinfo{volume}{71}},
  \bibinfo{pages}{020101} (\bibinfo{year}{2005}).

\bibitem[{\citenamefont{Ono et~al.}(2002)\citenamefont{Ono, O\char39{}Hern,
  Durian, Langer, Liu, and Nagel}}]{Ono02}
\bibinfo{author}{\bibfnamefont{I.~K.} \bibnamefont{Ono}},
  \bibinfo{author}{\bibfnamefont{C.~S.} \bibnamefont{O\char39{}Hern}},
  \bibinfo{author}{\bibfnamefont{D.~J.} \bibnamefont{Durian}},
  \bibinfo{author}{\bibfnamefont{S.~A.} \bibnamefont{Langer}},
  \bibinfo{author}{\bibfnamefont{A.~J.} \bibnamefont{Liu}}, \bibnamefont{and}
  \bibinfo{author}{\bibfnamefont{S.~R.} \bibnamefont{Nagel}},
  \bibinfo{journal}{Phys. Rev. Lett.} \textbf{\bibinfo{volume}{89}},
  \bibinfo{pages}{095703} (\bibinfo{year}{2002}).

\bibitem[{\citenamefont{Barrat}(2003)}]{Barrat02}
\bibinfo{author}{\bibfnamefont{J.-L.} \bibnamefont{Barrat}},
  \bibinfo{journal}{J. Phys.: Condens. Matter} \textbf{\bibinfo{volume}{15}},
  \bibinfo{pages}{S1} (\bibinfo{year}{2003}).

\bibitem[{\citenamefont{Barrat and Kob}(1999)}]{Barrat99}
\bibinfo{author}{\bibfnamefont{J.-L.} \bibnamefont{Barrat}} \bibnamefont{and}
  \bibinfo{author}{\bibfnamefont{W.}~\bibnamefont{Kob}},
  \bibinfo{journal}{Europhys. Lett.} \textbf{\bibinfo{volume}{46}},
  \bibinfo{pages}{637} (\bibinfo{year}{1999}).

\bibitem[{\citenamefont{Kob and Barrat}(1999)}]{Kob99}
\bibinfo{author}{\bibfnamefont{W.}~\bibnamefont{Kob}} \bibnamefont{and}
  \bibinfo{author}{\bibfnamefont{J.-L.} \bibnamefont{Barrat}},
  \bibinfo{journal}{Eur. Phys. J. B} \textbf{\bibinfo{volume}{13}},
  \bibinfo{pages}{319} (\bibinfo{year}{1999}).

\bibitem[{\citenamefont{Latz}(2000)}]{Latz00}
\bibinfo{author}{\bibfnamefont{A.}~\bibnamefont{Latz}}, \bibinfo{journal}{J.
  Phys.: Condens. Matter} \textbf{\bibinfo{volume}{12}}, \bibinfo{pages}{6353}
  (\bibinfo{year}{2000}).

\bibitem[{\citenamefont{Ilg and Barrat}(2007)}]{Ilg07}
\bibinfo{author}{\bibfnamefont{P.}~\bibnamefont{Ilg}} \bibnamefont{and}
  \bibinfo{author}{\bibfnamefont{J.-L.} \bibnamefont{Barrat}},
  \bibinfo{journal}{Europhys. Lett.} \textbf{\bibinfo{volume}{79}},
  \bibinfo{pages}{26001} (\bibinfo{year}{2007}).

\bibitem[{\citenamefont{Langer and Manning}(2007)}]{Langer07}
\bibinfo{author}{\bibfnamefont{J.~S.} \bibnamefont{Langer}} \bibnamefont{and}
  \bibinfo{author}{\bibfnamefont{M.~L.} \bibnamefont{Manning}},
  \bibinfo{journal}{Phys. Rev. E} \textbf{\bibinfo{volume}{76}},
  \bibinfo{pages}{056107} (\bibinfo{year}{2007}).

\bibitem[{\citenamefont{Greinert et~al.}(2006)\citenamefont{Greinert, Wood, and
  Bartlett}}]{Greinert06}
\bibinfo{author}{\bibfnamefont{N.}~\bibnamefont{Greinert}},
  \bibinfo{author}{\bibfnamefont{T.}~\bibnamefont{Wood}}, \bibnamefont{and}
  \bibinfo{author}{\bibfnamefont{P.}~\bibnamefont{Bartlett}},
  \bibinfo{journal}{Phys. Rev. Lett.} \textbf{\bibinfo{volume}{97}},
  \bibinfo{pages}{265702} (\bibinfo{year}{2006}).

\bibitem[{\citenamefont{Abou and Gallet}(2004)}]{Abou04}
\bibinfo{author}{\bibfnamefont{B.}~\bibnamefont{Abou}} \bibnamefont{and}
  \bibinfo{author}{\bibfnamefont{F.}~\bibnamefont{Gallet}},
  \bibinfo{journal}{Phys. Rev. Lett.} \textbf{\bibinfo{volume}{93}},
  \bibinfo{pages}{160603} (\bibinfo{year}{2004}).

\bibitem[{\citenamefont{Maggi et~al.}()\citenamefont{Maggi, d.~Leonardo, Dyre,
  and Ruocco}}]{Maggi09}
\bibinfo{author}{\bibfnamefont{C.}~\bibnamefont{Maggi}},
  \bibinfo{author}{\bibfnamefont{R.}~\bibnamefont{d.~Leonardo}},
  \bibinfo{author}{\bibfnamefont{J.~C.} \bibnamefont{Dyre}}, \bibnamefont{and}
  \bibinfo{author}{\bibfnamefont{G.}~\bibnamefont{Ruocco}},
  \bibinfo{note}{arXiv:0812.0740}.

\bibitem[{\citenamefont{{C.\,Godr\`{e}che\,and\,J.
  M.\,Luck}}(2000)}]{Godreche00a}
\bibinfo{author}{\bibnamefont{{C.\,Godr\`{e}che\,and\,J. M.\,Luck}}},
  \bibinfo{journal}{J. Phys. A: Math. Gen.} \textbf{\bibinfo{volume}{33}},
  \bibinfo{pages}{1\!1\!51} (\bibinfo{year}{2000}).

\bibitem[{\citenamefont{Godr\`{e}che and {J. M. Luck}}(2000)}]{Godreche00b}
\bibinfo{author}{\bibfnamefont{C.}~\bibnamefont{Godr\`{e}che}}
  \bibnamefont{and} \bibinfo{author}{\bibnamefont{{J. M. Luck}}},
  \bibinfo{journal}{J. Phys. A: Math. Gen.} \textbf{\bibinfo{volume}{33}},
  \bibinfo{pages}{9141} (\bibinfo{year}{2000}).

\bibitem[{\citenamefont{Calabrese and Gambassi}(2002)}]{Calabrese02}
\bibinfo{author}{\bibfnamefont{P.}~\bibnamefont{Calabrese}} \bibnamefont{and}
  \bibinfo{author}{\bibfnamefont{A.}~\bibnamefont{Gambassi}},
  \bibinfo{journal}{Phys. Rev. E} \textbf{\bibinfo{volume}{65}},
  \bibinfo{pages}{066120} (\bibinfo{year}{2002}).

\bibitem[{\citenamefont{Mayer et~al.}(2003)\citenamefont{Mayer, Berthier,
  Garrahan, and Sollich}}]{Mayer03}
\bibinfo{author}{\bibfnamefont{P.}~\bibnamefont{Mayer}},
  \bibinfo{author}{\bibfnamefont{L.}~\bibnamefont{Berthier}},
  \bibinfo{author}{\bibfnamefont{J.~P.} \bibnamefont{Garrahan}},
  \bibnamefont{and} \bibinfo{author}{\bibfnamefont{P.}~\bibnamefont{Sollich}},
  \bibinfo{journal}{Phys. Rev. E} \textbf{\bibinfo{volume}{68}},
  \bibinfo{pages}{016116} (\bibinfo{year}{2003}).

\bibitem[{\citenamefont{Sollich et~al.}(2002)\citenamefont{Sollich, Fielding,
  and Mayer}}]{Sollich02}
\bibinfo{author}{\bibfnamefont{P.}~\bibnamefont{Sollich}},
  \bibinfo{author}{\bibfnamefont{S.}~\bibnamefont{Fielding}}, \bibnamefont{and}
  \bibinfo{author}{\bibfnamefont{P.}~\bibnamefont{Mayer}}, \bibinfo{journal}{J.
  Phys.: Condens. Matter} \textbf{\bibinfo{volume}{14}}, \bibinfo{pages}{1683}
  (\bibinfo{year}{2002}).

\bibitem[{\citenamefont{{F. Corberi {\it et al.}}}(2003)}]{Corberi03}
\bibinfo{author}{\bibnamefont{{F. Corberi {\it et al.}}}}, \bibinfo{journal}{J.
  Phys A: Math. Gen.} \textbf{\bibinfo{volume}{36}}, \bibinfo{pages}{4729}
  (\bibinfo{year}{2003}).

\bibitem[{\citenamefont{Calabrese and Gambassi}(2004)}]{Calabrese04}
\bibinfo{author}{\bibfnamefont{P.}~\bibnamefont{Calabrese}} \bibnamefont{and}
  \bibinfo{author}{\bibfnamefont{A.}~\bibnamefont{Gambassi}},
  \bibinfo{journal}{J. Stat. Mech.: Theo. Exp.} p. \bibinfo{pages}{P07013}
  (\bibinfo{year}{2004}).

\bibitem[{\citenamefont{Calabrese and Gambassi}(2005)}]{Calabrese05}
\bibinfo{author}{\bibfnamefont{P.}~\bibnamefont{Calabrese}} \bibnamefont{and}
  \bibinfo{author}{\bibfnamefont{A.}~\bibnamefont{Gambassi}},
  \bibinfo{journal}{J. Phys. A: Math. Gen.} \textbf{\bibinfo{volume}{38}},
  \bibinfo{pages}{R133} (\bibinfo{year}{2005}).

\bibitem[{\citenamefont{Kr{\"u}ger and Fuchs}(2009)}]{Krueger09}
\bibinfo{author}{\bibfnamefont{M.}~\bibnamefont{Kr{\"u}ger}} \bibnamefont{and}
  \bibinfo{author}{\bibfnamefont{M.}~\bibnamefont{Fuchs}},
  \bibinfo{journal}{Phys. Rev. Lett.} \textbf{\bibinfo{volume}{102}},
  \bibinfo{pages}{135701} (\bibinfo{year}{2009}).

\bibitem[{\citenamefont{Fuchs and Cates}(2002)}]{Fuchs02}
\bibinfo{author}{\bibfnamefont{M.}~\bibnamefont{Fuchs}} \bibnamefont{and}
  \bibinfo{author}{\bibfnamefont{M.~E.} \bibnamefont{Cates}},
  \bibinfo{journal}{Phys. Rev. Lett.} \textbf{\bibinfo{volume}{89}}
  (\bibinfo{year}{2002}).

\bibitem[{\citenamefont{Fuchs and Cates}(2003)}]{Fuchs03}
\bibinfo{author}{\bibfnamefont{M.}~\bibnamefont{Fuchs}} \bibnamefont{and}
  \bibinfo{author}{\bibfnamefont{M.~E.} \bibnamefont{Cates}},
  \bibinfo{journal}{Faraday Discuss.} \textbf{\bibinfo{volume}{123}},
  \bibinfo{pages}{267} (\bibinfo{year}{2003}).

\bibitem[{\citenamefont{Fuchs and {M. E. Cates}}(2005)}]{Fuchs05}
\bibinfo{author}{\bibfnamefont{M.}~\bibnamefont{Fuchs}} \bibnamefont{and}
  \bibinfo{author}{\bibnamefont{{M. E. Cates}}}, \bibinfo{journal}{J. Phys.:
  Cond. Mat.} \textbf{\bibinfo{volume}{17}}, \bibinfo{pages}{1681}
  (\bibinfo{year}{2005}).

\bibitem[{\citenamefont{Fuchs and Cates}(2009)}]{Fuchs09}
\bibinfo{author}{\bibfnamefont{M.}~\bibnamefont{Fuchs}} \bibnamefont{and}
  \bibinfo{author}{\bibfnamefont{M.~E.} \bibnamefont{Cates}},
  \bibinfo{journal}{J. Rheol.} \textbf{\bibinfo{volume}{53}},
  \bibinfo{pages}{957} (\bibinfo{year}{2009}).

\bibitem[{\citenamefont{Fuchs}(2008)}]{Fuchs08b}
\bibinfo{author}{\bibfnamefont{M.}~\bibnamefont{Fuchs}},
  \bibinfo{journal}{Advances in Polymer Science}  (\bibinfo{year}{2008}),
  \bibinfo{note}{submitted, ArXiv:0810.2505}.

\bibitem[{\citenamefont{Besseling et~al.}(2007)\citenamefont{Besseling, Weeks,
  Schofield, and Poon}}]{Besseling07}
\bibinfo{author}{\bibfnamefont{R.}~\bibnamefont{Besseling}},
  \bibinfo{author}{\bibfnamefont{E.~R.} \bibnamefont{Weeks}},
  \bibinfo{author}{\bibfnamefont{A.~B.} \bibnamefont{Schofield}},
  \bibnamefont{and} \bibinfo{author}{\bibfnamefont{W.~C.~K.}
  \bibnamefont{Poon}}, \bibinfo{journal}{Phys. Rev. Lett.}
  \textbf{\bibinfo{volume}{99}}, \bibinfo{eid}{028301} (\bibinfo{year}{2007}).

\bibitem[{\citenamefont{Zausch et~al.}(2008)\citenamefont{Zausch, Horbach,
  Laurati, Egelhaaf, Brader, {Th.~Voigtmann}, and Fuchs}}]{Zausch08}
\bibinfo{author}{\bibfnamefont{J.}~\bibnamefont{Zausch}},
  \bibinfo{author}{\bibfnamefont{J.}~\bibnamefont{Horbach}},
  \bibinfo{author}{\bibfnamefont{M.}~\bibnamefont{Laurati}},
  \bibinfo{author}{\bibfnamefont{S.}~\bibnamefont{Egelhaaf}},
  \bibinfo{author}{\bibfnamefont{J.~M.} \bibnamefont{Brader}},
  \bibinfo{author}{\bibnamefont{{Th.~Voigtmann}}}, \bibnamefont{and}
  \bibinfo{author}{\bibfnamefont{M.}~\bibnamefont{Fuchs}}, \bibinfo{journal}{J.
  Phys.: Condens. Matter} \textbf{\bibinfo{volume}{20}},
  \bibinfo{pages}{404210} (\bibinfo{year}{2008}).

\bibitem[{Var()}]{Varnik08}
\bibinfo{note}{F. Varnik. {\it Complex Systems} ed M. Tokuyama and I. Oppenheim
  (Amer. Inst. of Physics, 2008) p 160}.

\bibitem[{\citenamefont{Dhont}(1996)}]{Dhont}
\bibinfo{author}{\bibfnamefont{J.~K.~G.} \bibnamefont{Dhont}},
  \emph{\bibinfo{title}{An Introduction to Dynamics of Colloids}}
  (\bibinfo{publisher}{Elsevier science}, \bibinfo{address}{Amsterdam},
  \bibinfo{year}{1996}).

\bibitem[{\citenamefont{Gazuz et~al.}(2009)\citenamefont{Gazuz, Puertas, {Th.
  Voigtmann}, and Fuchs}}]{Gazuz08}
\bibinfo{author}{\bibfnamefont{I.}~\bibnamefont{Gazuz}},
  \bibinfo{author}{\bibfnamefont{A.~M.} \bibnamefont{Puertas}},
  \bibinfo{author}{\bibnamefont{{Th. Voigtmann}}}, \bibnamefont{and}
  \bibinfo{author}{\bibfnamefont{M.}~\bibnamefont{Fuchs}},
  \bibinfo{journal}{Phys. Rev. Lett.} \textbf{\bibinfo{volume}{102}},
  \bibinfo{pages}{248302} (\bibinfo{year}{2009}).

\bibitem[{\citenamefont{Habdas et~al.}(2004)\citenamefont{Habdas, Schaar,
  Levitt, and Weeks}}]{Habdas04}
\bibinfo{author}{\bibfnamefont{P.}~\bibnamefont{Habdas}},
  \bibinfo{author}{\bibfnamefont{D.}~\bibnamefont{Schaar}},
  \bibinfo{author}{\bibfnamefont{A.~C.} \bibnamefont{Levitt}},
  \bibnamefont{and} \bibinfo{author}{\bibfnamefont{E.~R.} \bibnamefont{Weeks}},
  \bibinfo{journal}{Europhys. Lett.} \textbf{\bibinfo{volume}{67}},
  \bibinfo{pages}{477} (\bibinfo{year}{2004}).

\bibitem[{\citenamefont{Harada and i.~Sasa}(2005)}]{Harada05}
\bibinfo{author}{\bibfnamefont{T.}~\bibnamefont{Harada}} \bibnamefont{and}
  \bibinfo{author}{\bibfnamefont{S.}~\bibnamefont{i.~Sasa}},
  \bibinfo{journal}{Phys. Rev. Lett.} \textbf{\bibinfo{volume}{95}},
  \bibinfo{pages}{130602} (\bibinfo{year}{2005}).

\bibitem[{\citenamefont{Elrick}(1962)}]{Elrick62}
\bibinfo{author}{\bibfnamefont{E.~D.} \bibnamefont{Elrick}},
  \bibinfo{journal}{Austral. J. Phys.} \textbf{\bibinfo{volume}{15}},
  \bibinfo{pages}{283} (\bibinfo{year}{1962}).

\bibitem[{\citenamefont{Szamel}(2004)}]{Szamel}
\bibinfo{author}{\bibfnamefont{G.}~\bibnamefont{Szamel}},
  \bibinfo{journal}{Phys. Rev. Lett.} \textbf{\bibinfo{volume}{93}},
  \bibinfo{pages}{178301} (\bibinfo{year}{2004}).

\bibitem[{\citenamefont{Kr\"uger et~al.}()\citenamefont{Kr\"uger, Weysser,
  Zausch, Horbach, and Voigtmann}}]{twotime}
\bibinfo{author}{\bibfnamefont{M.}~\bibnamefont{Kr\"uger}},
  \bibinfo{author}{\bibfnamefont{F.}~\bibnamefont{Weysser}},
  \bibinfo{author}{\bibfnamefont{J.}~\bibnamefont{Zausch}},
  \bibinfo{author}{\bibfnamefont{J.}~\bibnamefont{Horbach}}, \bibnamefont{and}
  \bibinfo{author}{\bibfnamefont{T.}~\bibnamefont{Voigtmann}},
  \bibinfo{note}{in preparation}.

\bibitem[{\citenamefont{Kr{\"u}ger}(2009)}]{Krueger09b}
\bibinfo{author}{\bibfnamefont{M.}~\bibnamefont{Kr{\"u}ger}},
  \emph{\bibinfo{title}{Properties of Non-Equilibrium States: Dense Colloidal
  Suspensions under Steady Shearing}} (\bibinfo{publisher}{PhD Thesis},
  \bibinfo{address}{Universit{\"a}t Konstanz}, \bibinfo{year}{2009}),
  \urlprefix\url{http://nbn-resolving.de/urn:nbn:de:bsz:352-opus-80732}.

\bibitem[{\citenamefont{Hansen and McDonald}(1986)}]{Hansen}
\bibinfo{author}{\bibfnamefont{J.-P.} \bibnamefont{Hansen}} \bibnamefont{and}
  \bibinfo{author}{\bibfnamefont{I.~R.} \bibnamefont{McDonald}},
  \emph{\bibinfo{title}{Theory of Simple Liquids -- 2nd ed.}}
  (\bibinfo{publisher}{Academic press limited}, \bibinfo{address}{London},
  \bibinfo{year}{1986}).

\bibitem[{foo({\natexlab{a}})}]{footnote1}
\bibinfo{note}{The violating term according to Eq.~\eqref{eq:gl} reads
  $\Delta\chi_{\bf q}(t)=N_{\bf q}(0)\,\Phi_{\mathbf{q}}(0)t+\mathcal{O}(t^2)$.
  While $\Phi_{\mathbf{q}}(0)=1$, it follows with Eqs.~\eqref{eq:ch} and
  \eqref{eq:violap} that $N_{\bf q}(0)=\mathcal{O}(|\dot\gamma|)$.}

\bibitem[{\citenamefont{Varnik and Henrich}(2006)}]{Varnik06a}
\bibinfo{author}{\bibfnamefont{F.}~\bibnamefont{Varnik}} \bibnamefont{and}
  \bibinfo{author}{\bibfnamefont{O.}~\bibnamefont{Henrich}},
  \bibinfo{journal}{Phys. Rev. B} \textbf{\bibinfo{volume}{73}},
  \bibinfo{eid}{174209} (\bibinfo{year}{2006}).

\bibitem[{\citenamefont{G{\"o}tze}(1984)}]{Goetze84}
\bibinfo{author}{\bibfnamefont{W.}~\bibnamefont{G{\"o}tze}},
  \bibinfo{journal}{Z. Phys. B} \textbf{\bibinfo{volume}{56}},
  \bibinfo{pages}{139} (\bibinfo{year}{1984}).

\bibitem[{Goe()}]{Goetze91}
\bibinfo{note}{W. G{\"otze}. {\it Liquids, freezing and glass transition} ed
  J.-P. Hansen, D. Levesque and J. Zinn-Justin (Amsterdam, 1991) p 287}.

\bibitem[{\citenamefont{Crassous et~al.}(2008)\citenamefont{Crassous,
  Siebenb{\"u}rger, Ballauff, Drechsler, Hajnal, Henrich, and
  Fuchs}}]{Crassous08}
\bibinfo{author}{\bibfnamefont{J.~J.} \bibnamefont{Crassous}},
  \bibinfo{author}{\bibfnamefont{M.}~\bibnamefont{Siebenb{\"u}rger}},
  \bibinfo{author}{\bibfnamefont{M.}~\bibnamefont{Ballauff}},
  \bibinfo{author}{\bibfnamefont{M.}~\bibnamefont{Drechsler}},
  \bibinfo{author}{\bibfnamefont{D.}~\bibnamefont{Hajnal}},
  \bibinfo{author}{\bibfnamefont{O.}~\bibnamefont{Henrich}}, \bibnamefont{and}
  \bibinfo{author}{\bibfnamefont{M.}~\bibnamefont{Fuchs}}, \bibinfo{journal}{J.
  Chem. Phys.} \textbf{\bibinfo{volume}{128}}, \bibinfo{pages}{204902}
  (\bibinfo{year}{2008}).

\bibitem[{\citenamefont{Siebenb{\"u}rger
  et~al.}(2009)\citenamefont{Siebenb{\"u}rger, Fuchs, Winter, and
  Ballauff}}]{Siebenbuerger09}
\bibinfo{author}{\bibfnamefont{M.}~\bibnamefont{Siebenb{\"u}rger}},
  \bibinfo{author}{\bibfnamefont{M.}~\bibnamefont{Fuchs}},
  \bibinfo{author}{\bibfnamefont{H.}~\bibnamefont{Winter}}, \bibnamefont{and}
  \bibinfo{author}{\bibfnamefont{M.}~\bibnamefont{Ballauff}},
  \bibinfo{journal}{J. Rheol.} \textbf{\bibinfo{volume}{53}},
  \bibinfo{pages}{707} (\bibinfo{year}{2009}).

\bibitem[{foo({\natexlab{b}})}]{footnote2}
\bibinfo{note}{More sophisticated versions of Eq.~\eqref{eq:mod} exist
  \cite{Fuchs03,Fuchs09}, which are not necessary for our purpose.}

\bibitem[{\citenamefont{Varnik}(2006)}]{Varnik06b}
\bibinfo{author}{\bibfnamefont{F.}~\bibnamefont{Varnik}}, \bibinfo{journal}{J.
  Chem. Phys.} \textbf{\bibinfo{volume}{125}}, \bibinfo{eid}{164514}
  (\bibinfo{year}{2006}).

\bibitem[{\citenamefont{Graham}(1980)}]{graham}
\bibinfo{author}{\bibfnamefont{R.}~\bibnamefont{Graham}}, \bibinfo{journal}{Z.
  Physik B - Cond. Mat.} \textbf{\bibinfo{volume}{40}}, \bibinfo{pages}{149}
  (\bibinfo{year}{1980}).

\bibitem[{\citenamefont{McLennan}(1988)}]{McLennan}
\bibinfo{author}{\bibfnamefont{J.~A.} \bibnamefont{McLennan}},
  \emph{\bibinfo{title}{Introduction to Non-equilibrium Statistical Mechanics}}
  (\bibinfo{publisher}{Prentice Hall}, \bibinfo{address}{New York},
  \bibinfo{year}{1988}).

\bibitem[{\citenamefont{Baiesi et~al.}(2009)\citenamefont{Baiesi, Maes, and
  Wynants}}]{Baiesi09}
\bibinfo{author}{\bibfnamefont{M.}~\bibnamefont{Baiesi}},
  \bibinfo{author}{\bibfnamefont{C.}~\bibnamefont{Maes}}, \bibnamefont{and}
  \bibinfo{author}{\bibfnamefont{B.}~\bibnamefont{Wynants}},
  \bibinfo{journal}{Phys. Rev. Lett.} \textbf{\bibinfo{volume}{103}},
  \bibinfo{pages}{010602} (\bibinfo{year}{2009}).

\bibitem[{\citenamefont{Landau and Lifshitz}(1959)}]{Landauh}
\bibinfo{author}{\bibfnamefont{L.~D.} \bibnamefont{Landau}} \bibnamefont{and}
  \bibinfo{author}{\bibfnamefont{E.~M.} \bibnamefont{Lifshitz}},
  \emph{\bibinfo{title}{Course of Theoretical Physics, Volume 6: Fluid
  Mechanics}} (\bibinfo{publisher}{Pergamon Press}, \bibinfo{address}{Oxford},
  \bibinfo{year}{1959}).

\bibitem[{\citenamefont{Pine et~al.}(2005)\citenamefont{Pine, Gollub, Brady,
  and Leshansky}}]{Pine05}
\bibinfo{author}{\bibfnamefont{D.~J.} \bibnamefont{Pine}},
  \bibinfo{author}{\bibfnamefont{J.~P.} \bibnamefont{Gollub}},
  \bibinfo{author}{\bibfnamefont{J.~F.} \bibnamefont{Brady}}, \bibnamefont{and}
  \bibinfo{author}{\bibfnamefont{A.~M.} \bibnamefont{Leshansky}},
  \bibinfo{journal}{Nature} \textbf{\bibinfo{volume}{438}},
  \bibinfo{pages}{997} (\bibinfo{year}{2005}).

\bibitem[{\citenamefont{Gollub and Pine}(2006)}]{Gollub06}
\bibinfo{author}{\bibfnamefont{J.~P.} \bibnamefont{Gollub}} \bibnamefont{and}
  \bibinfo{author}{\bibfnamefont{D.~J.} \bibnamefont{Pine}},
  \bibinfo{journal}{Physics Today} \textbf{\bibinfo{volume}{59}},
  \bibinfo{pages}{8} (\bibinfo{year}{2006}).

\bibitem[{\citenamefont{Bouchaud}(1992)}]{Bouchaud92}
\bibinfo{author}{\bibfnamefont{J.~P.} \bibnamefont{Bouchaud}},
  \bibinfo{journal}{J. Phys. I} \textbf{\bibinfo{volume}{2}},
  \bibinfo{pages}{1705} (\bibinfo{year}{1992}).

\bibitem[{\citenamefont{Barrat and M{\'e}zard}(1995)}]{Barrat95}
\bibinfo{author}{\bibfnamefont{A.}~\bibnamefont{Barrat}} \bibnamefont{and}
  \bibinfo{author}{\bibfnamefont{M.}~\bibnamefont{M{\'e}zard}},
  \bibinfo{journal}{J. Phys. I} \textbf{\bibinfo{volume}{5}},
  \bibinfo{pages}{941} (\bibinfo{year}{1995}).

\bibitem[{\citenamefont{Monthus and Bouchaud}(1996)}]{Monthus96}
\bibinfo{author}{\bibfnamefont{C.}~\bibnamefont{Monthus}} \bibnamefont{and}
  \bibinfo{author}{\bibfnamefont{J.~P.} \bibnamefont{Bouchaud}},
  \bibinfo{journal}{J. Phys. A} \textbf{\bibinfo{volume}{29}},
  \bibinfo{pages}{3847} (\bibinfo{year}{1996}).

\bibitem[{\citenamefont{Rinn et~al.}(2000)\citenamefont{Rinn, Maass, and
  Bouchaud}}]{Rinn00}
\bibinfo{author}{\bibfnamefont{B.}~\bibnamefont{Rinn}},
  \bibinfo{author}{\bibfnamefont{P.}~\bibnamefont{Maass}}, \bibnamefont{and}
  \bibinfo{author}{\bibfnamefont{J.~P.} \bibnamefont{Bouchaud}},
  \bibinfo{journal}{Phys. Rev. Lett.} \textbf{\bibinfo{volume}{84}},
  \bibinfo{pages}{5403} (\bibinfo{year}{2000}).

\bibitem[{\citenamefont{Fielding and Sollich}(2002)}]{Fielding02}
\bibinfo{author}{\bibfnamefont{S.}~\bibnamefont{Fielding}} \bibnamefont{and}
  \bibinfo{author}{\bibfnamefont{P.}~\bibnamefont{Sollich}},
  \bibinfo{journal}{Phys. Rev. Lett.} \textbf{\bibinfo{volume}{88}},
  \bibinfo{pages}{050603} (\bibinfo{year}{2002}).

\bibitem[{\citenamefont{Fielding et~al.}(2000)\citenamefont{Fielding, Sollich,
  and Cates}}]{Fielding00}
\bibinfo{author}{\bibfnamefont{S.~M.} \bibnamefont{Fielding}},
  \bibinfo{author}{\bibfnamefont{P.}~\bibnamefont{Sollich}}, \bibnamefont{and}
  \bibinfo{author}{\bibfnamefont{M.~E.} \bibnamefont{Cates}},
  \bibinfo{journal}{J. Rheol.} \textbf{\bibinfo{volume}{44}},
  \bibinfo{pages}{323} (\bibinfo{year}{2000}).

\bibitem[{\citenamefont{Gardiner}(1985)}]{Gardiner}
\bibinfo{author}{\bibfnamefont{C.~W.} \bibnamefont{Gardiner}},
  \emph{\bibinfo{title}{Handbook of Stochastic Methods}}
  (\bibinfo{publisher}{Springer}, \bibinfo{address}{Berlin},
  \bibinfo{year}{1985}).

\bibitem[{foo({\natexlab{c}})}]{footnote3}
\bibinfo{note}{We ignore the fact that the movement is discontinuous.}

\bibitem[{\citenamefont{Kr{\"u}ger and Fuchs}()}]{Krueger09d}
\bibinfo{author}{\bibfnamefont{M.}~\bibnamefont{Kr{\"u}ger}} \bibnamefont{and}
  \bibinfo{author}{\bibfnamefont{M.}~\bibnamefont{Fuchs}}, \bibinfo{note}{in
  preparation}.

\bibitem[{\citenamefont{Speck and Seifert}(2009)}]{Speck09}
\bibinfo{author}{\bibfnamefont{T.}~\bibnamefont{Speck}} \bibnamefont{and}
  \bibinfo{author}{\bibfnamefont{U.}~\bibnamefont{Seifert}},
  \bibinfo{journal}{Phys. Rev. E.} \textbf{\bibinfo{volume}{79}},
  \bibinfo{pages}{040102(R)} (\bibinfo{year}{2009}).

\bibitem[{\citenamefont{G\"otze and Latz}(1989)}]{Goetze89}
\bibinfo{author}{\bibfnamefont{W.}~\bibnamefont{G\"otze}} \bibnamefont{and}
  \bibinfo{author}{\bibfnamefont{A.}~\bibnamefont{Latz}}, \bibinfo{journal}{J.
  Phys.:Condens. Matter} \textbf{\bibinfo{volume}{1}}, \bibinfo{pages}{4169}
  (\bibinfo{year}{1989}).

\bibitem[{\citenamefont{Fuchs and Kroy}(2002)}]{Fuchs02b}
\bibinfo{author}{\bibfnamefont{M.}~\bibnamefont{Fuchs}} \bibnamefont{and}
  \bibinfo{author}{\bibfnamefont{K.}~\bibnamefont{Kroy}}, \bibinfo{journal}{J.
  Phys.:Condens. Matter} \textbf{\bibinfo{volume}{14}}, \bibinfo{pages}{9223}
  (\bibinfo{year}{2002}).

\bibitem[{\citenamefont{Cichocki and Hess}(1987)}]{Cichocki87}
\bibinfo{author}{\bibfnamefont{B.}~\bibnamefont{Cichocki}} \bibnamefont{and}
  \bibinfo{author}{\bibfnamefont{W.}~\bibnamefont{Hess}},
  \bibinfo{journal}{Physica A} \textbf{\bibinfo{volume}{141}},
  \bibinfo{pages}{475} (\bibinfo{year}{1987}).

\bibitem[{\citenamefont{Fuchs et~al.}(1998)\citenamefont{Fuchs, G{\"o}tze, and
  Mayr}}]{Fuchs98}
\bibinfo{author}{\bibfnamefont{M.}~\bibnamefont{Fuchs}},
  \bibinfo{author}{\bibfnamefont{W.}~\bibnamefont{G{\"o}tze}},
  \bibnamefont{and} \bibinfo{author}{\bibfnamefont{M.~R.} \bibnamefont{Mayr}},
  \bibinfo{journal}{Phys. Rev. E} \textbf{\bibinfo{volume}{58}},
  \bibinfo{pages}{3384} (\bibinfo{year}{1998}).

\bibitem[{\citenamefont{Henrich et~al.}(2007)\citenamefont{Henrich, Pfeifroth,
  and Fuchs}}]{Henrich07}
\bibinfo{author}{\bibfnamefont{O.}~\bibnamefont{Henrich}},
  \bibinfo{author}{\bibfnamefont{O.}~\bibnamefont{Pfeifroth}},
  \bibnamefont{and} \bibinfo{author}{\bibfnamefont{M.}~\bibnamefont{Fuchs}},
  \bibinfo{journal}{J. Phys: Condens. Matter} \textbf{\bibinfo{volume}{19}},
  \bibinfo{pages}{205132} (\bibinfo{year}{2007}).

\bibitem[{\citenamefont{{M. Kr{\"u}ger {\it et al.}}}()}]{incoherent}
\bibinfo{author}{\bibnamefont{{M. Kr{\"u}ger {\it et al.}}}}, \bibinfo{note}{in
  preparation}.

\end{thebibliography}
\end{document}